\documentclass{tcibook}
\usepackage{fancyhea}
\usepackage{work}
\usepackage{bm}       
\usepackage{graphicx}
\usepackage{hyperref}      
\usepackage{multirow}

\newcommand{\nc}{\newcommand}  

\def\beq{\begin{equation}}
\def\eeq#1{\label{#1}\end{equation}}
\def\eeqn{\end{equation}}

\newenvironment{Eqnarray}%
   {\arraycolsep 0.14em\begin{eqnarray}}{\end{eqnarray}}
\def\beqa{\begin{Eqnarray}}
\def\eeqa#1{\label{#1}\end{Eqnarray}}
\def\eeqan{\end{Eqnarray}}

\nc{\ra}{\rightarrow}  
\nc{\slsh}{\slash\hspace*{-0.22cm}}
\def\Re{{\cal R \mskip-4mu \lower.1ex \hbox{\it e}\,}}
\def\Im{{\cal I \mskip-5mu \lower.1ex \hbox{\it m}\,}}

\nc{\vev}[1]{ \left\langle {#1} \right\rangle }
\nc{\bra}[1]{ \langle {#1} | }
\nc{\ket}[1]{ | {#1} \rangle }
\nc{\fb}{\,{\rm fb}^{-1}}
\nc{\ev}{{\rm eV}}
\nc{\kev}{{\rm keV}}
\nc{\Mev}{{\rm MeV}}
\nc{\gev}{{\rm GeV}}
\nc{\tev}{{\rm TeV}}
\nc{\mev}{{\rm MeV}}

\def\del{\partial}
\def\Dslash{\not{\hbox{\kern-4pt $D$}}}
\def\dslash{\not{\hbox{\kern-2pt $\del$}}}
\def\pslash{\not{\hbox{\kern-2pt $p$}}}
\def\ETmiss{ \not{\hbox{\kern-4pt $E$}}_T }

\def\msb{{\bar{\ssstyle M \kern -1pt S}}}

\begin{document}

\def\bibname{References}
\bibliographystyle{plain}

\raggedbottom

\pagenumbering{roman}

\parindent=0pt
\parskip=8pt
\setlength{\evensidemargin}{0pt}
\setlength{\oddsidemargin}{0pt}
\setlength{\marginparsep}{0.0in}
\setlength{\marginparwidth}{0.0in}
\marginparpush=0pt


\pagenumbering{arabic}

\renewcommand{\chapname}{chap:intro_}
\renewcommand{\chapterdir}{.}
\renewcommand{\arraystretch}{1.25}
\addtolength{\arraycolsep}{-3pt}

\thispagestyle{empty}
\begin{centering}
\vfill

{\Huge\bf Planning the Future of U.S. Particle Physics}

{\Large \bf Report of the 2013 Community Summer Study}

\vfill

{\Huge \bf Chapter 8: Instrumentation Frontier}

\vspace*{2.0cm}
{\Large \bf Conveners: M. Demarteau, R. Lipton, H. Nicholson, and I. Shipsey}
\pagenumbering{roman}

\vfill

{\large  Study Conveners: M. Bardeen, W. Barletta, L.~A.~T.~Bauerdick, R. Brock,
D.~Cronin-Hennessy, M.~Demarteau, M.~Dine, J.~L. Feng, M. Gilchriese,
S. Gottlieb, J.~L.~Hewett, R. Lipton, H.~Nicholson, M.~E. Peskin,
S. Ritz, I.~Shipsey, H. Weerts}\\
\vspace{1cm}

{\large Division of Particles and Fields Officers in 2013:
J.~L. Rosner (chair), 
I. Shipsey (chair-elect), 
N. Hadley (vice-chair),
P. Ramond (past chair)}\\
\vspace{1cm}

{\large Editorial Committee:
R.~H. Bernstein,
N. Graf,
P. McBride,
M.~E. Peskin,
J.~L. Rosner,
N.~Varelas,
K. Yurkewicz}

\vfill

\end{centering}

\newpage
\mbox{\null}

\vspace{3.0cm}

{\Large \bf Authors of Chapter 8:}

\vspace{2.0cm}
 {\bf M. Demarteau, R. Lipton, H. Nicholson, I. Shipsey}, D.~Akerib,
 A.~Albayrak-Yetkin, J.~Alexander, J.~Anderson, M.~Artuso, D.~Asner,
 R.~Ball, M.~Battaglia, C.~Bebek, J.~Beene, Y.~Benhammou,
 E.~Bentefour, M.~Bergevin, A.~Bernstein, B.~Bilki, E.~Blucher,
 G.~Bolla, D.~Bortoletto, N.~Bowden, G.~Brooijmans, K.~Byrum,
 B.~Cabrera,  G.~Cancelo, J.~Carlstrom, B.~Casey, C.~Chang,
 J.~Chapman, C.H.~Chen, I.~Childres, D.~Christian, M.~Convery,
 W.~Cooper J.~Corso, J.~Cumalat, P.~Cushman, C.~Da~Via, S.~Dazeley,
 P.~Debbins, G.~Deptuch, S.~Dhawan, V.~Di~Benedetto, B.~DiGiovene,
 Z.~Djurcic,
 S.~Dye, A.~Elagin, J.~Estrada, H.~Evans, E.~Etzion, J.~Fast,
 C.~Ferretti, P.~Fisher,
 B.~Fleming, K.~Francis, P.~Friedman, H.~Frisch, M.~Garcia-Sciveres,
 C.~Gatto, G.~Geronim,
 G.~Gilchriese, S.~Golwala,
 C.~Grant, A.~Grillo, E.~Gr\"unendahl, P.~Gorham, L.~Guan,
 G.~Gutierrez, C.~Haber, J.~Hall, G.~Haller, C.~Hast,
 U.~Heintz, T.~Hemmick,
 D.~G.~Hitlin, C.~Hogan, M.~Hohlmann, E.~Hoppe, L.~Hsu, M.~Huffer,
 K.~Irwin, F.~Izraelevitch, G.~Jennings, M.~Johnson, A.~Jung,
 H.~Kagan, C.~Kenney, S.~Kettell,
 R.~Khanna, V.~Khristenko, F.~Krennrich, K.~Kuehn,  R.~Kutschke,
 J.~Learned, A. T.~Lee, D.~Levin, T.~Liu, A. T. K.~Liu,
 D.~Lissauer, J.~Love,  D.~Lynn, D.~MacFarlane, S.~Magill,
 S.~Majewski, J.~Mans, J.~Maricic, P.~Marleau, A.~Mazzacane, D.~McKinsey,
 J.~Mehl, A.~Mestvirisvilli, S.~Meyer,
 N.~Mokhov, M.~Moshe, A.~Mukherjee, P.~Murat,  S.~Nahn,
 M.~Narain, P.~Nadel-Turonski, M.~Newcomer,
 K.~Nishimura, D.~Nygren, E.~Oberla, Y.~Onel, M.~Oreglia, J.~Orrell,
 J.~Paley, A.~Para, S.~Parker, V.~Polychronakos, S.~Pordes,
 P.~Privitera, A.~Prosser, M.~Pyle, J.~Raaf,  E.~Ramberg, R.~Rameika,
 B.~Rebel, J.~Repond, D.~Reyna, L.~Ristori, R.~Rivera,
 A.~Ronzhin, R.~Rusack, J.~Russ, A.~Ryd, H.~Sadrozinski, H.~Sahoo,
 M. C.~Sanchez, C.~Sanzeni, S.~Schnetzer,
 S.~Seidel, A.~Seiden, I.~Schmidt, A.~Shenai, T.~Shutt, Y.~Silver,
 W.~Smith, D.~Snowden-Ifft,
 A.~Sonnenschein, D.~Southwick, L.~Spiegel, M.~Stanitzki,
 S.~Striganov, D.~Su, R.~Sumner, R.~Svoboda,
 M.~Sweany,
 R.~Talaga,
 R.~Tayloe, S.~Tentindo, N.~Terentiev, J.~Thom-Levy, C.~Thorn,
 J.~Tiffenberg, W.~Trischuk, R.~Tschirhart, M.~Turner, D.~Underwood,
 L.~Uplegger, J.~Urheim,
 M.~Vagins, K.~Van~Bibber, G.~Varner, R.~Varner, J.~Va'vra, H.~Von der
 Lippe, R.~Wagner, S.~Wagner,  C.~Weaverdyck,  H.~Wenzel, A.~Weinstein,
 M.~Wetstein, A.~White, R.~Wigman, P.~Wilson, D.~Winn, P.~Winter,
 C.~Woody, L.~Xia, J. Q.~Xie, Z.~Ye, M. F.~Yeh, T.~Yetkin, J.~H.~Yoo,
 J.~Yu, J. M.~Yu, S.~Zeller, J. L.~Zhang, J.J.~Zhu, B.~Zhou, R. Y.~Zhu, B.~Zitzer

 \tableofcontents

\newpage

\mbox{\null}

\newpage

\pagenumbering{arabic}


\pagenumbering{arabic}

\renewcommand{\chapname}{chap:intro_}
\renewcommand{\chapterdir}{.}
\renewcommand{\arraystretch}{1.25}
\addtolength{\arraycolsep}{-3pt}


\setcounter{chapter}{7}     
\chapter{Instrumentation Frontier}
\label{chap:instrum}

\begin{center}\begin{boldmath}


\begin{large} {\bf Conveners: M. Demarteau, R. Lipton, H. Nicholson, I. Shipsey}\end{large}

D.~Akerib, 
A.~Albayrak-Yetkin,
J.~Alexander,
J.~Anderson, 
M.~Artuso,
D.~Asner
% ...

\end{boldmath}\end{center}


\section{Introduction}
Innovation in instrumentation is central to discovery science.   
Progress in understanding 
the laws which govern the most fundamental building blocks of nature has been achieved 
largely through technological advances in instrumentation that have made increasingly 
sophisticated experiments and analysis of the data from these experiments possible.  The 
field of high-energy physics (HEP) is viewed as an incubator of innovation, which has led to fundamental 
discoveries in particle physics as well as breakthroughs in other fields of science.\footnote{High-energy
physics and particle physics are used interchangably in this Chapter.} In many cases, novel physics 
experiments have been enabled by advances in technology,\footnote{To quote 
Freeman Dyson: ``New directions in science are launched by new tools much more often than by new concepts.  
The effect of a concept-driven revolution is to explain old things in new ways.  The effect of a tool-driven 
revolution is to discover new things that have to be explained~\cite{Dyson}.''}
whereas in others, physics needs have driven necessary technological development.

The physics requirements of many experiments in the Energy, Intensity and Cosmic Frontiers 
entail very large-scale detectors that use technologies that cannot be scaled up without 
the cost becoming prohibitive.  The very large costs involved in building 
and operating many existing HEP detectors, combined with increasingly tight budgetary constraints, 
make it difficult for the HEP community to provide substantial investment in new technologies, many of
which have the potential to make significant contributions to the future HEP program but which carry 
a substantial risk of failure.

Newer, riskier technologies will almost certainly need to be developed to 
enable study of particle interactions with high statistics at higher energies, to look for 
extremely rare decays or ultra-low cross section processes, and to obtain and analyze increasingly large data 
samples.  The need to develop new, transformational technologies has been recognized in recent government panel 
recommendations that are promoting the need for riskier but higher reward research~\cite{PCAST}. 
Investment in the development of new technologies is an investment in the sustainability of the field. 
It has therefore become very important for the 
HEP community to establish a mechanism for developing and implementing a coherent vision 
for the future of {\it generic} HEP instrumentation development. 

The Snowmass Instrumentation Frontier is an outgrowth of a process initiated by an American
Physical Society Division of Particles and Fields (DPF) task force established in 2010.  This 
task force was charged with addressing the needs, organization, and health of instrumentation 
in U.S. particle physics.  The major task force recommendation was that the DPF form a standing 
committee to address a number of outstanding issues related to instrumentation~\cite{DPF-TF}:

\begin{itemize}
\item
Make available up-to-date information on elements of the national detector R\&D program.
\item
Expand coordination among the national laboratories, and provide a forum (and information) 
for enhanced access to selected resources at the national laboratories for university groups.
\item
Stimulate detector R\&D through workshops and studies in a coordinated way.
\item
Stimulate new ideas, especially those of a scale that requires substantial collaboration,
particularly between laboratories and universities.
\item
Instigate a concerted effort to involve industry in workshops and studies with the
intent of later involvement in R\&D.
\item
Coordinate joint educational activities at schools and other community events.
\item
Explore opportunities for attracting and providing careers for high-energy physicists 
in instrumentation. 
\item
Act as a resource for the funding agencies by helping them establish SBIR categories,
improve the response to program solicitations, and investigate other opportunities.
\end{itemize}

In response to this recommendation the DPF established the Coordinating Panel for Advanced Detectors 
(CPAD). This committee, supplemented by liaisons to the physics frontiers 
and technologies, was the basis of the organization of the Instrumentation Frontier and will 
continue its work after the Snowmass process is complete.

There are at least two time scales to be considered in formulating a national vision for 
instrumentation: (1) the intermediate term ($<$ 10 years), where the 
community has a reasonable understanding of the physics goals, and (2) the longer term, 
where technologies being developed now will ultimately become the enablers of future experiments  
which are either technologically inaccessible or prohibitively costly today.  We must be able to 
effectively communicate these needs to the funding agencies.


This report is organized as follows. 
Sections~\ref{sec:instrum-energy}, \ref{sec:instrum-intensity} and 
\ref{sec:instrum-cosmic}
describe major experiments with significant U.S. involvement in the Energy, 
Intensity, and Cosmic Frontiers, and the technologies currently 
employed by these experiments. We identify existing technologies that need to be refined and 
new technologies that need to be developed in order to address the current or near term needs 
of these experiments.  
When possible, we also describe projected needs beyond the next decade.
In section~\ref{sec:instrum-tech}
we describe several widely-used, key technologies  
identified during the Snowmass process, in which future R\&D would enable the U.S. HEP
community to maintain leadership in detector 
development in a global HEP community and build a strong, future experimental research program.    
Section~\ref{sec:instrum-facilities} lists the existing facilities, and 
partnerships available for instrumentation are described in section~\ref{sec:instrum-universities}. 
In section~\ref{sec:instrum-leader}
we describe issues which must be addressed to maintain detector R\&D leadership in the future.  
Finally, in section~\ref{sec:instrum-conclusions}
we provide some conclusions based on the material presented in the previous sections. 
 
\section{The Energy Frontier}
\label{sec:instrum-energy}

\subsection{Hadron colliders}
\subsubsection{Current and planned projects}

At this time, the Energy Frontier is embodied by the Large Hadron Collider (LHC) 
and the two large detectors ATLAS and CMS.  The primary physics goals of the 
LHC detectors in the next 5 years are (1) to measure branching ratios and related 
couplings of the recently discovered Higgs boson and to 
characterize its properties in as much detail as possible to determine whether 
or not it is the Standard Model Higgs, (2) to search for physics beyond the 
Standard Model including supersymmetry, dark matter, and extra dimensions, and 
(3) to carry out a measurement of $WW$ scattering.  This implies maintaining detector 
performance both at low-energy scales for Higgs studies and the highest scales 
in searches for new physics.  The LHC will resume running in 2014 at 13 or 14~TeV 
center of mass energy with an average instantaneous luminosity of $5 \times 10^{33}$~cm$^{-2}$s$^{-1}$, 
which is already beyond the design parameters of the experiments.

ATLAS and CMS are both large multipurpose detectors.  CMS is more compact, with a 4T 
central field.  It includes an all-silicon tracker, a lead-tungstate electromagnetic 
calorimeter, a  brass/scintillator sampling hadron calorimeter, and a muon system with 
drift tubes, resistive plate chambers, and cathode strip chambers. The CMS compact 
design and high magnetic field enables the use of high-resolution crystal electromagnetic 
calorimetry and all-silicon tracking~\cite{CMS:2006}. 

ATLAS has a lower, 2T, field with an inner silicon tracking section, an outer straw 
tube transition radiation tracker, and a liquid argon sampling electromagnetic calorimeter 
with an outer scintillator-tile hadronic calorimeter.  The ATLAS muon system utilizes 
air core toroid magnets with both barrel tracking chambers and tracking chambers in an 
inner and outer wheel configuration at each end. The ATLAS liquid argon calorimeter
is radiation-hard and provides excellent pattern recognition. The ATLAS muon system 
provides an independent measurement of muon momenta~\cite{ATLAS:1999uwa}.

Both the CMS and ATLAS experiments performed very well in the initial physics runs. Both experiments 
have devised operations, trigger, and analysis strategies to cope with high occupancies caused 
by the 50 ns bunch-crossing interval and high instantaneous luminosity.  The experiments 
have also identified significant physics limitations.  Both trackers are more massive than 
optimal. The CMS endcap crystals are beginning to suffer from radiation damage.  Both pixel 
detectors will suffer losses due to buffer overflows at higher luminosity.  Triggers 
for both experiments are close to saturation.

Both ATLAS and CMS are currently being upgraded~\cite{Brooijmans:2013gba}. 
These Phase-I upgrades include improvements to the trigger and DAQ systems and replacement of the inner 
pixel detectors~\cite{Smith:2013oka}.   The new detectors will utilize low mass mechanics and 
CO$_2$ cooling to reduce tracker mass.   Hybrid photodiodes are being replaced by silicon 
photomultipliers, increasing the granularity of the CMS hadron calorimeter. These changes will 
enable continued operation of these detectors to integrated luminosities of 300 fb$^{-1}$, 
expected to be reached by the year 2022. 

\subsubsection{Future outlook and needs}
The European Strategy calls for an upgrade of the LHC instantaneous luminosity to 
$5 \times 10^{34}$~cm$^{-2}$s$^{-1}$ (HL-LHC).  ATLAS and CMS will need to operate at an average of 
$\sim 140$ interactions per 25 nsec beam crossing time and expect to obtain 3000 fb$^{-1}$ of 
integrated luminosity.  To achieve their physics goals the experiments will need to maintain 
at least the current values of resolution for tracking and calorimetry and maintain the ability to 
selectively trigger on specific physics events at reasonable thresholds. 

The projected increase of the instantaneous luminosity at the HL-LHC by an order of magnitude
will lead to nonlinear increases in raw trigger rates. Improved rejection power will be needed
at each trigger level. Both CMS and ATLAS plan to utilize the increasing bandwidth, storage, and 
processing capabilities of modern electronics to enable robust and flexible triggering.
Plans include increases of the level 1 (L1) and level 2 (L2) trigger bandwidth and an increased 
L1 trigger latency. In addition, both CMS and ATLAS foresee access to the full granularity of the 
detector at L1, which enables increased sophistication of 
trigger algorithms, and a L1 track trigger utilizing either on-detector correlations or 
regions of interest.  All of this will need low-power, high-speed, radiation-hard optical data links.

Measurements of $WW$ scattering and Higgs self-coupling require efficient identification of jets at 
high pseudorapidity $\eta$.\footnote{Pseudorapidity is defined as $\eta \equiv -\ln(\tan(\frac{\theta}{2}))$ 
where $\theta$ is the usual polar angle from the beam direction.}
This is an enormous challenge, with singles rates on the order of $5 \times 10^{9}$ Hz at 
the inner radius of a forward calorimeter. This will require high-rate, large-area, radiation-hard 
detectors.
Ideally, a calorimeter would provide $\sim 100$ ps time tags to separate primary vertices, 
allow for dual readout of electromagnetic 
and hadronic energies, and provide for detailed pattern recognition.

Current detectors and electronics used for inner vertex layers are not sufficiently radiation-hard and
will either have difficulty or will be unable to handle projected luminosity increases.  Further 
R\&D on 3D, diamond, or thin silicon sensors is necessary for layers at small radii.  Silicon will 
have to be kept near -20$^{\circ}C$ to avoid reverse annealing and to control the leakage currents.  Similarly, 
inner layers will need radiation-hard electronics with acceptable single-event upset rates and good 
performance after exposure to doses approaching a Grad.

The most powerful new tools envisioned for the HL-LHC upgrades are the L1 track triggers. An L1 track
trigger, which includes the ability to identify primary vertex $z$ location, can eliminate much of the pileup 
background. This requires a pixelated tracker with pixels that can correlate information 
with other detectors.  In the CMS track trigger design, track stubs greater than 2 GeV can be 
found by using hit correlations between 
silicon sensors separated by $\approx$1-4~mm. This requires unprecedented connectivity and intelligence 
in the tracker electronics.  The  design of the high-speed, complex, on-detector electronics is 
in turn directly limited by the power that can be delivered by DC-DC converters or serial power. The 
increased pixelization of the tracker and increasing data bandwidths will continue to increase power 
density and cooling demands. New support structure and cooling technologies will be needed to 
control mass and therefore preserve the resolution of the trackers and calorimeters.

\subsubsection{Beyond the next decade} 

Discoveries at the Large Hadron Collider (LHC) may point to hadron machines with energies beyond 
lepton collider capabilities, such as a Very Large Hadron Collider (VLHC). Experiments at these facilities 
indicate clear instrumentation needs.  New physics signals from additional Higgs bosons, supersymmetric
particles, or other unexpected particles will likely be increasingly buried in QCD 
background.  This points to the need for experiments in high-luminosity environments with selective triggers 
and robust data acquisition bandwidth. This in turn implies increased pixelization in the tracking detectors 
and calorimeters to enable experiments to identify the correct primary interaction and reduce total hit occupancy.  
Multiple interaction backgrounds can also be mitigated by picosecond-level timing or by limiting the region 
of the interaction diamond where the shower of particles originated. Almost all system components  must be 
radiation-hard.

Advances in electronics and sensor technology can start to address these issues.  
Novel integrated-circuit packaging technologies hold the promise for large-area sensor front-end readout 
electronics bonding at very fine pitch, which could be considerably less expensive than conventional bump bonding.   
 Sensors can be designed to have lower capacitance, which will decrease noise, increase speed, and 
decrease power consumption.
Silicon sensors can be operated in avalanche or Geiger mode, substantially increasing speed 
while eliminating the need for high currents in front-end amplifiers. Light sensors can also be pixelated 
using Silicon PhotoMultiplier (SiPM) or Large Area Picosecond Photodetector (LAPPD)-style technologies.  
Power can be reduced towards 10 pJ/bit and bandwidth increased in on-detector data transmission utilizing 
new radiation-hard optical transmission technologies or silicon photonics. New semiconductors 
are being developed for commercial DC-DC conversion which may provide both radiation hardness and increased 
voltage step-down ratios at high efficiency. Advances in opto-photonics may lead to completely novel 
detectors. 

\subsection{Lepton colliders}
\subsubsection{Current and planned projects}
Lepton colliders appear to be the machines of choice for precision measurements of the properties of the 
Higgs boson and Standard Model particles, to elucidate the nature of supersymmetry, extra dimensions, or 
other new physics discovered at the LHC.  The candidates for $e^+e^-$ colliders are an International Linear 
Collider (ILC) which can be tuned for a center-of-mass energy between 200-500 GeV, 
with a potential energy limit of 1 TeV; CLIC, a 
higher energy but less well-developed $e^+e^-$ collider; and TLEP, a Higgs factory in a $\approx$ 80 km 
ring~\cite{ILC:TDRv3, CLIC:LCS2012, TLEP:2013}.  
An alternative is a muon collider, which has the advantage of being both much more compact, because of the lower 
radiation losses of the muons, and being scalable to higher energy.  

One application for an $e^+e^-$ collider is  measuring the properties of the Higgs 
boson, including its couplings to fermions and other bosons. 
Furthermore, an $e^+e^-$ collider allows a precise determination of the properties of the top 
quark, and study of possible signals of new physics through the electroweak production of new particles 
and through signals of these interactions in $W$, $Z$, and two-fermion processes.  $e^+e^-$ collider experiments 
will be sensitive to supersymmetric partners of known particles, new heavy gauge bosons, 
extra spatial dimensions, and particles connected with strongly-coupled theories of electroweak symmetry 
breaking.   

There has been considerable design work carried 
out for the silicon detector (SiD) and the International Large Detector (ILD), which 
are designed for the ILC but are also being used as a basis for possible CLIC detectors, where 
beam backgrounds will be substantially greater.  
SiD is a general purpose detector with (1) a multilayer 
barrel and endcap silicon 
pixel vertex detector, (2) a multilayer barrel and endcap 
silicon strip tracker,
(3) a multilayer electromagnetic calorimeter consisting 
of thinner inner and thicker outer layers of sheets of tungsten absorber interleaved with silicon 
wafers with hexagonal 12~mm$^2$ silicon pixels, (4) a highly segmented hadronic calorimeter, 
(5) a 5T solenoidal magnet, and (6) an iron 
flux return instrumented with either scintillator or resistive plate chambers.  ILD is less compact than SiD 
and features a large TPC for tracking. Both detectors are designed to be used as a complete system for 
Particle Flow Analysis (PFA) to measure jet energies, charged leptons, photons, 
and missing energy~\cite{ILC:TDRv4}.  
CLIC has higher energy, more beamstrahlung, and 0.5~ns  bunch spacing and will require faster detectors and 
electronics as well as thicker calorimetry~\cite{CLIC:CDR2012}.

\subsubsection{Key issues and next steps}
The SiD and ILD detector design concepts for the ILC are fairly mature.   
Key issues that will need to be addressed and solved for 
ILC detectors are: fabrication of low mass vertex detectors and trackers, power delivery and low mass supports, 
and the design and affordable implementation of precise particle flow or dual readout calorimetry.

Both the ILC and CLIC aspire to 
measure the energy of hadron showers with an unprecedented resolution that can separate hadronic $W$ and $Z$ decays.  
The particle flow technique, which aims at a di-jet mass resolution of  
better than $5\%$, uses highly pixelated calorimeters to identify energy deposited by 
neutral hadrons and adds that energy to the energy of tracks in a jet.  An alternative dual readout technique,
which shows promising results in a collimated test beam and also aims to achieve better than $5\%$ jet energy 
resolution, attempts to separately measure hadronic and electromagnetic components of a jet and correct the jet 
energy for losses due to nuclear breakup.  

There is an increasing focus on understanding the fundamental limitations to hadronic calorimeter energy 
resolution to enable optimum resolution calorimeter designs in future detectors.  The CALICE collaboration has 
studied particle flow calorimetry for ILC and has developed test hadron calorimeters which have been extensively 
tested in the Fermilab and CERN test beams.  The CALICE digital hadron calorimeter concept could be the basis for 
experiments needing highly granulated charged particle calorimeters~\cite{CALICE:2012, DHC:2012}.
R\&D on dual readout calorimetry is being carried out by the CERN RD52 collaboration and others. Crystals 
capable of high-resolution total-absorption calorimetry are also being explored~\cite{DualReadout:2013}. 

\subsubsection{Beyond the next decade}
LHC measurements have not yet found any evidence for new physics. If new physics does exist, a multi-TeV 
lepton collider may provide a path to fully explore the spectrum.  $e^+e^-$ colliders have barriers, 
however, to scaling to the multi-TeV range.   ILC technology will be too expensive. Power requirements may 
place an upper limit on affordable CLIC energies. 
TLEP cannot reach center of mass energies beyond 350 GeV in an 80 km ring.  

These barriers to scaling do not apply to a muon collider.  A compact circular $\mu^{+}\mu^{-}$ collider 
can be built with acceptable energy loss and cost 
if the appropriate cooling and magnet technologies can be developed.  However, substantial experimental 
difficulties must be overcome to deal with the radiation background from muon decays in flight.  Decay 
backgrounds originate primarily from electromagnetic showers in absorbers and beamline components.  The 
overwhelming majority of these backgrounds is composed of low-energy particles, 
which produce signals at different times than
signals produced from particles from the interaction vertex.  The key to background rejection is fast, 
nanosecond-level timing in the tracker and calorimeters. Pixelization is needed both for resolution and 
to limit background occupancy. 
Experiments at a muon collider will combine the radiation hardness needs from the LHC, 
timing requirements of CLIC, and precision of experiments at the ILC.

\section{The Intensity Frontier}
\label{sec:instrum-intensity}

\subsection{The neutrino sector}
\subsubsection{Current and planned projects}

There are a number of unanswered questions about neutrino properties and about the relationship between three 
fundamental neutrino eigenstates $\nu_1$, $\nu_2$, $\nu_3$, and the mix of them in the physical neutrinos $\nu_e$, 
$\nu_{\mu}$, and $\nu_{\tau}$.  In particular, it is currently unclear whether neutrinos are Majorana or Dirac, 
whether or not the masses of $\nu_1$ and $\nu_2$ are greater or less than the mass of $\nu_3$, and whether neutrino 
interactions violate CP.  In addition, the absolute masses of the neutrinos are unknown. 

Reactor neutrino experiments include Daya Bay, which is currently running with eight detectors, and Double 
Chooz, which is running with one (far) detector out of a planned two detectors~\cite{Band:2012rk, Abe:2012tg}.  
These experiments are explicitly designed to measure the mixing angle $\theta_{13}$.  


Neutrinoless double-beta decay experiments are designed to determine if the neutrino is Majorana 
(its own anti-particle) or Dirac, since 
neutrinoless double-beta decay is impossible unless the neutrino is Majorana.  In addition, if the mass hierarchy is 
known to be inverted, the absence of a signal below a critical mass provides definite proof that the neutrinos 
must be Dirac.  Nuclear physics non-reactor experiments planned for the near future include three neutrinoless 
double-beta decay experiments, --- SNO+, Majorana, and Cuore --- and one additional experiment, Katrin.  
Katrin is designed to measure, or put a more stringent limit on, the neutrino mass by carefully measuring the 
tritium beta decay phase space endpoint.

\subsubsection{Key issues and next steps}

Large-mass, high-resolution, cost-effective detectors will be needed for future neutrino and proton decay 
experiments~\cite{Kettell}. 
To improve the analysis capabilities of neutrino detectors, the field has decided to move away from the 
workhorse water Cherenkov detector to liquid argon (LAr) tracking detectors, 
which have a 3D space point resolution of less than 1 mm 
and energy resolution of less than 10~MeV.  Although relatively small LAr Time Projection Chambers (TPCs) 
have been constructed and are in use 
in a variety of HEP experiments, considerable detector R\&D will be required to insure 
the required purity of very large 
volumes of liquid argon,\footnote{Electronegative contaminants in LAr must be reduced to 100 parts per trillion of 
oxygen equivalent contamination.} to generate and distribute the high voltages required for drifting ions in the 
chamber,\footnote{Voltage requirements are $\sim$500 Volts/meter over several meters.} to make the cryostats and 
cryogenic systems as inexpensive as possible, and to fabricate low noise (possibly cold) readout electronics for 
very large-volume LAr detectors, that will work reliably for tens of years.  It will also be desirable to provide a 
magnetic field in LAr chambers to help with particle identification.  In addition, a number of safety 
issues need to be resolved before a very large LAr detector can be deployed underground.

A number of technical issues need to be resolved in designing larger LAr detectors.  
Electronics may be warm or cold. The optimum electronics sampling time for a given detector needs to be 
determined.\footnote{Current and planned values are 0.197 $\mu$s for ArgoNeuT, 0.4 $\mu$s for ICARUS, and 0.5 $\mu$s 
for both MicroBooNE and LBNE.}  
Scintillation light readout may use photomultipliers, which require 
wavelength shifting of the 128 nm LAr scintillation light and which take up 
space,\footnote{These are used by MicroBooNE and ICARUS.} wavelength shifter paddles,\footnote{These are planned for 
LAr1 and LBNE.} or silicon photomultipliers, which are expensive but have a low profile.  Finally, liquid 
and gas two-phase argon detectors are being considered as possible candidates for future large neutrino detectors.
    
The cost of liquid scintillator is a problem for very large volume liquid scintillator detectors, and the cost of 
photomultipliers is a problem for both large volume liquid scintillator and water Cherenkov detectors.  A brighter 
and optically cleaner scintillator for future long baseline (JUNO (Daya Bay~II) and RENO-50) and future short 
baseline (SBL) reactor experiments is also highly desirable.\footnote{Short baseline reactor experiments have 
baselines less than 100 m.}  

A water-based liquid scintillator, capable of detecting the $p \rightarrow K^{+} \overline{\nu}$ channel 
that is invisible 
to a Cherenkov detector, has recently been developed at Brookhaven National Laboratory. It is about 
an order of magnitude 
cheaper than mineral oil-based liquid scintillator, and might be used to increase the sensitive area of 
the NO$\nu$A far 
detector.   In addition, new metallic loading methods in water-based scintillators provide metallic ions, 
which can customize the properties of future scintillation detectors~\cite{MinFang}
in novel ways.  For example, the SNO+ experiment has taken advantage of this new technology, with Te-doped 
scintillator as a projected neutrinoless double-beta decay source and detector.  

The LAPPD program is developing fast, high-spatial-resolution photodetectors which may allow the 
development of an optical TPC using fast timing and good photon spatial resolution for tracking 
and the different characteristics of Cherenkov and scintillation light for particle identification.  If large-volume, 
low-noise detectors can be developed, detecting low-energy reactor antineutrinos using coherent neutrino scattering 
with the corresponding two to three orders of magnitude increase in interaction cross section becomes very attractive.  In 
addition, neutron tagging techniques could help with flavor separation in large water Cherenkov detectors such as the 
proposed Japanese Hyper Kamiokande detector.\footnote{Gadolinium doping would allow detection of neutron captures in water 
via Gd(n,$\gamma$) which gives an 8 MeV gamma cascade versus a 2.2 MeV gamma from H(n,$\gamma$).}  

Advanced water Cherenkov detectors are also being used by other parts of the scientific community.  The WATCHMAN collaboration 
of universities and national laboratories for nuclear non-proliferation is conducting a site search for a kiloton-scale 
advanced water detector to passively obtain detailed information on the operation of commercial nuclear reactors.  

Plans are being made to scale up the 0$\nu\beta\beta$
decay Stanford/SLAC EXO-200 experiment to a 5 ton experiment. The main 
technological concerns are maintaining the required xenon purity for the larger TPC, 
identifying large-area low-radioactivity 
ultraviolet optical detectors to replace the large area avalanche photodiodes used in EXO-200, and developing
cold, front-end electronics with acceptably low levels of radioactivity.  
Another possibility is to detect barium atoms
that are produced when Xe undergoes either two-neutrino or neutrinoless double-beta decay.  
If a next-generation xenon 0$\nu\beta\beta$  
decay experiment can develop the technology for tagging barium atoms with high efficiency, the experiment will be 
virtually free of backgrounds from non-xenon decays.  
In addition, if future xenon double-beta decay experiments use high-pressure 
xenon gas instead of liquid xenon, there is the possibility of an order of 
magnitude improvement in energy resolution 
due to the much lower Fano factor in xenon gas.  Better energy resolution would help remove two-neutrino backgrounds 
(and any remaining non-double-beta decay backgrounds) 
which are not under the 0$\nu\beta\beta$ decay energy peak.

Fermilab is researching the possible use of solid $^{136}$Xe as a 0$\nu\beta\beta$ detector.  A 2.3 MeV energy deposition 
generates $\sim$10$^7$ phonon quanta, which would predict an energy resolution of 0.3\% at 1\% collection efficiency.  
This is about one order of magnitude better energy resolution than
that of EXO-200.  Further R\&D on the actual properties of sub-Kelvin crystalline xenon is
necessary because few studies have been done under these conditions.  In addition, there is a need to grow large 
transparent xenon crystals and read out signals from large crystals.

The U.S. is currently in a position to take a leadership role in all the areas identified in this section.
Fermilab is interested in developing a high power source of neutrinos, and a neutrino factory could be 
a useful source of intense neutrino beams in order to carry out high statistics neutrino experiments.

\subsubsection{Beyond the next decade}

Double-beta decay experiments benefit greatly from isotopic enrichment of the double-beta decay isotope.  It would be 
very beneficial to develop less expensive ways to enrich these isotopes.  In addition, liquid argon detectors 
used in rare decay experiments would be substantially improved if the ${}^{39}$Ar contamination, which is produced
by cosmic rays, could be removed inexpensively from ${}^{40}$Ar. 

Optical TPCs using water-based scintillators and inexpensive, fast, high-resolution photodetectors could be  
enlarged at considerably less cost than cryogenic, noble gas TPCs.  Ultra-low-threshold neutrino detectors have the 
potential for measuring low energy neutrino interactions with a coherent neutrino-nucleus cross section which is larger 
than a neutrino-nucleon interaction by a factor of roughly the square of the number of nucleons in the nucleus.

There is also a growing need to identify and stockpile construction materials that are free from radioactive 
contamination to reduce backgrounds in very large detector systems and double-beta decay experiments. 

\subsection{Rare decays}
\subsubsection{Current and planned projects}

There are currently two rare decay experiments running, the MEG experiment \cite{Nishiguchi:2011zz} searching for 
the lepton flavor-violating reaction $\mu \rightarrow e \gamma$, and KOTO \cite{Iwai:2012qya} which will attempt to 
measure the CP-violating reaction $K_L \rightarrow \pi^0 \nu \overline{\nu}$ with enough precision to look 
for new physics if there is a significant deviation from the Standard Model 
calculation.  Two other rare decay experiments are proposed: Mu2e, which will 
search for the lepton flavor changing decay $\mu \rightarrow e$ \cite{Knoepfel:2013ouy}, 
and ORKA, which will search for significant deviations from the Standard Model calculation of the decay 
$K^+ \rightarrow \pi^+ \nu \overline{\nu}$ \cite{Worcester:2012rd}.  

The Muon $g-2$ experiment will make precise measurements of the gyromagnetic ratio of the muon to explore with greater 
precision the disagreement between previous results and QED calculations~\cite{g-2}.

Currently, MEG has published a limit of 2.4$\times 10^{-12}$ on the branching ratio $\mu \rightarrow e \gamma$, and 
there is a proposed upgrade to a branching ratio measurement sensitivity of 6$\times 10^{-13}$.  
Key detector requirements for this upgrade consist of improving the rate capability of all detectors and improving 
energy, angular, and timing resolutions for the positron and photon arms of the detector.  A more highly granulated 
positron tracker is envisioned, along with a highly segmented, fast timing counter array using silicon photomultipliers.  
The photon arm, which uses the largest liquid xenon detector in the world, will be improved by replacing photomultiplier 
tubes with arrays of smaller photosensors, such as silicon or multi-anode photomultipliers. 
This will increase the granularity at the incident face.

KOTO is being constructed at J-PARC.  Since there are no charged particles in the reaction of interest, the 
challenges are to produce a neutron-free, pencil-like beam of $K_L$ mesons and to measure the energy and time of  
the photons from the $\pi^0$ decay.  This experiment needs a highly efficient, 
finely granulated gamma detector which completely surrounds the decay region. 

The Mu2e experiment is currently undergoing the DOE's critical decision (CD) process for final 
approval and is currently scheduled to begin running in fiscal year 2019.  The detector consists of a 
highly segmented straw tube tracker and a LYSO crystal calorimeter located in the warm bore of a solenoid
with a 1 Tesla magnetic field.  It is surrounded by a passive cosmic ray veto as well as an active veto using 
wavelength-shifting fibers embedded in scintillating material. The veto scintillator fibers will be read out by 
silicon photomultipliers.

The ORKA experiment will use an upgraded detector modeled on the Brookhaven E787 and E949 experiments.  The 
new experiment will increase the sensitivity by a factor of 100 over previous experiments, 
with a 10 times more intense
kaon beam and a 10 times more sensitive detector. 

\subsubsection{Key issues and next steps}

All the rare decay experiments benefit from highly segmented trackers and timing counters to reduce 
backgrounds~\cite{Kettell}.   
Currently, silicon photomultiplers are becoming the photodetectors of choice because of their high photon detection 
efficiency and high segmentation. The principal limitation is cost in covering large areas.
The LAPPD project could potentially provide photodetectors that could deliver precise timing in rare decay 
experiments.
 
\subsubsection{Beyond the next decade}
Project X will enable an extensive program of critical tests of the Standard Model 
in both rare muon and kaon processes. The flagship process is the CP-violating kaon decay 
$K_L \rightarrow \pi^0 \nu \overline{\nu}$, where a thousand-event experiment may stress the Standard Model 
understanding of flavor.  Simultaneously, lepton-flavor-violation processes involving charge exchanges will reach 
unprecedented sensitivities, with the exact program determined by progress in the current round of 
$\mu \rightarrow e \gamma$ and $\mu \rightarrow e$  conversion experiments.
These experiments will take advantage of the tremendous flux and exquisite time structure of the Project X 
beams~\cite{Kettell}. 
To fully capitalize on these accelerator capabilities, associated improvements in detector rate capability and 
timing resolution will be needed. These experiments will need fast, high resolution electromagnetic calorimetry; 
low mass, high-resolution, fast tracking; cost-effective, high efficiency photon detection; and high throughput 
streaming data acquisition systems. The  $K_L \rightarrow \pi^0 \nu \overline{\nu}$ calorimeter will need good photon pointing. 
These experiments would benefit from less expensive, highly segmented photodetectors with high quantum efficiency and immunity 
to magnetic fields.

\subsection{Heavy flavor factories}
\subsubsection{Current and planned projects}

Heavy flavor detectors are designed to make precision measurements of $B$, $B_s$, and $D$ mesons. The physics 
program includes measurements of their decay and mixing, the study of CP 
violation and measurement of the CKM angles, and a search for new physics in decays 
prohibited by the Standard Model or in branching ratios that differ significantly from Standard Model predictions.  
Three detectors, each with different capabilities, are carrying out precision heavy flavor measurements.
The LHCb detector~\cite{LHCb:UpgradeTDR} will search for new physics signatures in beauty and charm decays
with capabilities that extend to electroweak and top physics and studies of exotic states.
The Belle-II detector~\cite{Abe:2010sj} will search for  new CP-violating phases, flavor-changing 
neutral currents (FCNC) beyond the Standard Model and other manifestations of new physics in 
rare $B$ decays.
BES III~\cite{Wang:2007sm} is designed to do precision studies of $D$ decays and mixing, 
to look for new charmed states, and to provide precision data which can test 
lattice QCD calculations.   The BaBar detector is no longer taking data, but analysis of its data set is 
ongoing and, in 2012, it uncovered direct evidence for T violation in the $B^0$ system.

\subsubsection{Key issues and next steps}

Major upgrades are planned for the LHCb and Belle detectors. 
An upgrade to the LHCb detector has been approved for running at higher LHC luminosities.  
The LHCb collaboration is considering a triggerless data acquisition (DAQ) system running at 40~MHz for which 
new, radiation-hardened DAQ boards and high-speed optical links are required.  The experiment will
need a more highly segmented, radiation-hardened tracking system, a low-mass cooling system, and better 
capabilities for forming RF shielding foils.  
The upgrade of the Belle to Belle-II detector consists of installing a new two-layer DEPFET-based pixel 
detector, complemented by a four-layer silicon tracking system using double-sided silicon sensors. 
Quartz-based time-of-propagation counters will be used for particle identification in the barrel, 
read out with waveform sampling electronics. The trigger, data acquisition system, and readout 
electronics will also undergo a major upgrade. 
No upgrades are currently being planned for BES III.

The key issues that must be addressed in heavy flavor detector upgrades are almost entirely 
related to dealing with higher 
luminosities: LHCb in the range of $10^{34}$~cm$^{-2}$s$^{-1}$ and Belle II in the range of 
$10^{36}$~cm$^{-2}$s$^{-1}$~\cite{Kettell}.  Radiation hardness of electronics and 
vertex detector elements must be improved, faster electronics is needed, 
and granularity of tracking detectors must be 
increased to reduce the ambiguity due to increased occupancy. Better triggering and faster 
data acquisition systems must 
also be designed.  The addition of fast waveform sampling electronics will improve background 
rejection and particle identification. 

\subsubsection{Beyond the next decade}

There is another charm/tau factory under consideration, the BINP Super c/$\tau$ Factory with 
projected beam energies from 
1.0 to 2.5 GeV, a peak luminosity of $10^{35}$ at 2 GeV, and longitudinally polarized electrons at 
the interaction point~\cite{Roney:2013}.  This is one of six projects approved for further 
development by the Russian Governmental Commission 
on Innovations and High Technologies.

The United States was planning to be a major player in a Super-BaBar detector in Italy, 
using recycled parts from the original SLAC BaBar Detector.  
Now that that project has been canceled by the Italian government, the focus has shifted 
towards the BELLE-II and LHCb upgrades. The U.S. currently carries a major responsibility 
for the design, construction and readout of the quartz-based particle identification detector 
for the BELLE-II detector. 

\section{The Cosmic Frontier}
\label{sec:instrum-cosmic}
 
\subsection{Ultra-high-energy cosmic rays and neutrinos}
\subsubsection{Current and planned projects}

There are a number of experiments designed to measure the energy spectrum of ultra-high-energy 
cosmic rays and neutrinos, to determine their composition and source directionality, and to 
observe and study the Greisen-Zatsepin-Kuzmin (GZK) cutoff.\footnote{The GZK cutoff occurs when 
protons with energies greater than $5 \times 10^{19}$~eV interact with cosmic microwave background 
photons to produce a $\Delta^+$ which decays into $p, \pi^0$ or $n \pi^+$.}  

The Pierre Auger Observatory uses an array of 1600 water 
Cherenkov detectors, spaced 1.5~km apart, and 24 fluorescence detectors to study and characterize 
extensive air showers resulting from ultra-high-energy cosmic rays~\cite{Allekotte:2007sf, Abraham:2009pm}.  
The fluorescence-light-detecting telescopes 
record fluorescence in the ultraviolet spectrum from shower particle excitation of nitrogen.

The Ice Cube neutrino observatory at the South Pole~\cite{Kiryluk:2008dm} consists of a surface air 
shower detector and regularly spaced arrays of 5160 photomultiplier tubes (PMTs) buried in the South Pole 
ice between 1450 and 2450 meters below the surface. The PMTs record Cherenkov radiation from charged 
particles from neutrino interactions in the ice.  This experiment aims to better understand  
the origins of ultra-high-energy cosmic rays, search for cold dark matter particles with masses 
approaching a TeV, search for supersymmetric particles, and study neutrino oscillations over 
megaparsec baselines.  The PINGU upgrade consists of adding more densely packed photomultiplier  
strings to reduce the neutrino oscillation energy threshold and study matter effects in neutrino 
oscillations~\cite{PINGU:2013}.

The ANtarctic Impulsive Transient Antenna (ANITA) experiment is also designed to study ultra-high-energy 
cosmic neutrinos~\cite{ANITA:2013}.   It uses sensitive microwave antennae to detect electromagnetic 
showers produced by ultra-high-energy neutrinos in the Antarctic ice through the Askaryan 
effect.\footnote{The Askaryan effect is coherent radio emission from excess negative charge in an 
electromagnetic shower.}  The experiment attempts to locate the source of ultra-high-energy cosmic rays, 
measure model independent weak interaction cross sections, and look for new physics in neutrino flavor 
oscillations.

\subsubsection{Key issues and next steps}

The major limitation of experiments that aim to measure the spectrum of ultra-high-energy cosmic rays and 
neutrinos is the very low flux of high-energy particles from space.\footnote{A detector the size of Auger 
detects $\sim$30 cosmic rays with an energy of $\sim 10^{20}$~eV per year.}  Thus, it is very important to have 
high detection efficiency.  The Pierre Auger Observatory would greatly benefit from low-cost photodetectors 
with high spectral efficiency between 300 and 400~nm and the capability to survive high background currents.

The ANITA experiment would benefit from reducing the levels of temperature equivalent noise from 70~K to 
10~K in affordable amplifiers.\footnote{Amplifiers with noise levels of 70~K were made affordable 
because of the recent revolution in wireless technology.}  The experiment would also benefit from 
improvements in antenna technology, lower-power waveform digitizers with better dynamic range and 
throughput, and more flexible, addressable sample windows.  Continued R\&D on ASIC development for 
waveform capture is also very desirable. 

\subsection{Dark energy}
\subsubsection{Current and planned projects}

The discovery of cosmic acceleration, or dark energy, has spawned a new generation of experiments whose 
purpose is to elucidate this phenomenon.  These experiments fall into two classes: (1) distance measurements,  
to track the expansion of the universe, and (2) measurements of the growth of cosmic structure, 
to determine whether the acceleration 
is due to a cosmological dark energy  or modified gravity.  Distance measurements either use 
supernovae as standard candles or baryon acoustic oscillations to compare the length scale of galaxy 
clusters today to the clustering at the time of matter recombination.  Structure measurements either use  
weak lensing or galaxy cluster counting. 

The currently running Dark Energy Survey (DES) uses the 4~meter Blanco telescope with a 570 megapixel charge
coupled device (CCD) imager~\cite{DES:2005}.  The Baryon Oscillation Spectroscopic Survey (BOSS) is conducting a 
five year survey utilizing the Sloan Digital Sky Survey 2.5 meter telescope and a spectrometer, which was upgraded 
in 2009~\cite{Dawson:2012va}.  The proposed Mid-Scale Dark Energy Spectroscopic Instrument (MS-DESI), is scheduled 
to start running around 2018 and will consist of a 4~meter telescope coupled to 5000 robotically controlled fibers.  
It will obtain spectra for about 10\% of the available galaxies with a magnitude limit of 22.5 and a redshift of 
about 3.5~\cite{Levi:2013}.  Euclid, a joint European-NASA space imager and spectrograph scheduled to launch in 2020, 
will consist of a 1.2 meter telescope with a visible imager (VIS), near infrared (NIR) photometry and 
spectroscopy~\cite{Euclid:2012}.  The Large Synoptic Survey Telescope (LSST), an imager scheduled to 
begin taking data in 2021, will consist of an 8 meter telescope using six filters for visible photometry~\cite{LSST:2011}.  
It will have a magnitude limit of $\sim$27.5, will cover 20,000 deg$^2$ in 10 years, and measure $\sim$20 billion 
galaxies.  A key issue for the LSST experiment will be acquiring, managing, and analyzing huge data sets.

\subsubsection{Key issues and next steps}

All dark energy experiments should be able to perform high-resolution 
($R = \lambda / \Delta\lambda \geq$ 1000)
spectral analysis.  DES and LSST use multiple filters and large CCD cameras to carry out a 5 or 6 point spectral 
analysis.  The ability to carry out high-resolution spectral analysis with the camera itself would be 
a major technological step forward in this research.

There is a current R\&D effort underway to develop a larger spectrometer using Microwave Kinetic Inductance 
Detectors (MKIDs).  These are superconducting resonators.  When a photon strikes the superconductor, it breaks 
Cooper pairs, which changes the inductance of the resonator, which in turn changes the resonant frequency and 
its phase. The phase change is then measured, providing a time and energy measurement for each photon.  
The theoretical resolution $R$ is $\sim$100, and up to 2000 resonators could be multiplexed to the same readout line.

A proposed low resolution MKID spectrometer, Giga-Z  will consist of 100,000 MKIDs~\cite{Gigaz:2013}.
The built-in spectral resolution of the MKIDs allows the use of direct lens-coupled, aperture mask spectroscopy 
as opposed to the fiber-fed spectroscopy. This promises a far more efficient use of telescope time with wavelength 
coverage extending to the near infrared, which is important for observations of high redshift galaxies.
Outstanding issues are getting $R$ closer to the theoretical limit, improving the packaging and RF 
electronics, improving the quantum efficiency, and developing a geometrically flat RF cable that works at 
100 milliKelvin. 

\subsection{Gamma rays}
\subsubsection{Current and planned projects}

There are currently two principal techniques used for detecting atmospheric gamma rays in the range from tens 
of GeV up to $\sim$200 TeV: the atmospheric air Cherenkov technique and the water Cherenkov technique.  
The atmospheric air Cherenkov technique records images of air showers by focusing Cherenkov light onto
a telescope consisting of an array of photosensors ($\sim$1000 pixels) and a large optical reflector.
The VERITAS, H.E.S.S, and MAGIC observatories use this technique with multiple telescopes in coincidence to 
form stereoscopic views of air showers~\cite{VERITAS:1999, VERITAS:2011, HESS:2003, MAGIC:2005}.  
Existing Imaging Atmospheric Cherenkov Telescopes (IACTs) used in these observatories reconstruct the arrival 
direction with an accuracy of $<$0.1 degree and measure the gamma ray energy with a resolution of 10-15\% for 
cosmic rays with energies from 50 GeV to 50 TeV.   

The water Cherenkov technique has a wider field of view and higher duty cycle than the IACTs, but has less 
sensitivity and poorer angular resolution.  The shower particles reaching the water tanks produce 
Cherenkov light and are detected with photomultipliers.  These signals are used to reconstruct the arrival 
direction and energy of the air shower.  The HAWC (High Altitude Water Cherenkov) experiment will consist of 
an array of 300 water tanks with the ability to observe 50\% of the sky in a 24-hour period~\cite{HAWC:2013}.

The larger Cherenkov Telescope Array (CTA), with $\sim$100 telescopes, is currently in the planning 
stage~\cite{CTA:2009}.  It will provide multiple stereo views in shower reconstruction and should be sensitive 
to energies between 30 GeV and 300 TeV.  

These experiments can provide sensitive searches for gamma rays resulting from WIMP dark matter annihilation. 
They can also indirectly search for axion like particles (ALPs), which couple to gammas, by searching 
for a gamma-axion oscillation, which produces an anomalous transparency for gammas traversing the galaxies.
In addition, they can look for violations of Lorentz invariance by searching for energy-dependent photon speeds using 
pulsar timing sources and flares from distant blazars. 

\subsubsection{Key issues and next steps}

The Cherenkov telescope experiments have been using photomultiplier tubes at the focal point of large optical 
mirror dishes.  Increased-granularity (10,000 - 100,000 pixels) and high-efficiency large area photodetectors 
with immunity to sudden bright flashes of light are particularly desirable for this application.  The recent 
introduction of novel optical instrument designs, such as dual mirror systems, allows a large reduction in the 
size of the Cherenkov light images, enabling the use of arrays of smaller photodetectors such as silicon 
photomultipliers and multi-anode phototubes.  These devices, combined with inexpensive and high-density 
readout electronics with GHz sampling capability, will reduce cost and improve performance of detector 
arrays.  

For the CTA experiment, which is envisioned to have a telescope array covering from 1 to 3 km$^2$, a precision 
($<$ 1 nsec) clock distribution system with custom chips for array processing is essential for triggering. 
A wireless data acquisition system would be very useful for recording the data.

The HAWC experiment would benefit from high-efficiency, large-area photodetectors to instrument its 300 water 
tanks.  Precise clock signal distribution in trigger systems and data acquisition systems are also a challenge. 

\subsection{Heavy dark matter - WIMPs}
\subsubsection{Current and planned projects}

There are currently several experiments running and planned whose motivation is to directly detect weakly 
interacting massive particle (WIMP) dark matter by observing collisions of cosmic WIMPs with target nuclei.  
These experiments fall into six principal categories:

(1) Solid state detectors with transition edge phonon sensors to measure both the ionization and thermal 
energies of potential dark matter recoils with nuclei in the detector (CDMS~\cite{CDMS:2005})

(2) Solid state germanium detectors, which use shaped internal electric fields and very small 
capacitance readout contacts to reduce noise so that low-energy nuclear recoils at a level of about 500 eV 
can be detected (CoGeNT~\cite{CoGeNT:2012})

(3) Experiments looking for low mass WIMPs in Charge Coupled Devices (CCDs)
(DAMIC~\cite{DAMIC:2012})\footnote{The DArk Matter In CCD experiment, DAMIC, uses 10 250~$\mu$m thick CCDs with a ten 
$\mu$m pixel size cooled to -150 $^{\circ}$C with a total mass of 0.01 kg enclosed in a low-radiation package in a copper 
box to look for WIMPS in the 1 - 10 GeV mass range, where other direct search dark matter experiments do not have 
good sensitivity.}

(4) Two-stage liquid/gas xenon TPCs which detect ionization light (S1) from the nuclear recoil in 
the liquid and electro-luminescence light (S2) from electrons drifting through the gas and use their ratio to 
distinguish nuclear recoils from electron backgrounds (Xenon 100~\cite{Xenon100:2012}, LUX~\cite{LUX:2012})

5) Two-stage liquid/gas argon detectors (the Darkside family of detectors~\cite{Darkside:2011}),
which use S1 and S2 light as well as S1 pulse shape discrimination, and single stage liquid argon detectors 
(the DEAP/CLEAN family of detectors~\cite{MiniCLEAN}), which use pulse shape discrimination alone to distinguish 
nuclear recoils in the liquid argon from electron recoils resulting from natural radioactivity at the level of 
better than a part in 10$^8$

(6) Bubble chambers, which adjust their operating pressure to nucleate a single bubble from a nuclear recoil 
(COUPP)~\cite{COUPP:2008}.   


\subsubsection{Key issues and next steps}

Since the WIMP mass is not known, it is important to increase the sensitivity of dark matter experiments to cover 
the broadest possible mass range, in particular the region from 1 MeV to 1 GeV.  Because the nuclear recoil energies 
are very small for low mass WIMPS, viable detectors must be sensitive at or near the single electron-ion pair limit.
One technique for reducing the WIMP mass detection threshold is to increase the phonon signal in a CDMS 
semiconductor 
dark matter detector by operating it at high voltage bias and using the electric potential energy in the detector 
to produce Luke-Neganov phonons as the electron-ion pairs traverse the detector, in addition to the phonons produced 
by the recoil itself.  This has the effect of discriminating nuclear recoil signals for low WIMP masses from 
background electron recoils.  Neutron backgrounds and small target mass are the most significant problems for DAMIC, 
and better shielding is required to improve sensitivity.  The experiment has moved from the Fermilab Minos tunnel to 
SNOLab.  


  
Most of the large scale experiments described in section 8.4.4.1 are planning upgrades to greater detector mass.  
CDMS is planning to upgrade from 9.3~kg of active germanium detector mass using 3'' diameter, 1'' thick 
detectors in the Soudan underground laboratory to 100-200 kg of active germanium detector mass using 3.9'' 
diameter, 1.3'' thick detectors in SNOLab.  In the much longer term, a CDMS detector with 1 metric ton of active 
germanium detector mass using 6'' diameter, 2'' thick detectors operating either at SNOLab or the Sanford Lab 
is anticipated.

Both xenon experiments, Xenon100 and LUX, are planning to upgrade.  Xenon100 plans to upgrade from 100~kg of 
xenon detector mass to 1 ton (Xenon1T)~\cite{XENON1T:2012}.  LUX plans to upgrade from 350~kg of xenon with a 
fiducial mass of 100~kg to a 7 ton xenon detector with a fiducial mass of 5 tons (LZ)~\cite{LZ:2011}.   
R\&D will be necessary to develop the high-voltage, low-radioactivity feedthroughs and high-voltage components to provide 
the necessary drift field.  Both experiments also need to develop xenon purification systems, particularly to remove 
radon and radioactive krypton from xenon.  The LUX/LZ experiment uses low-radioactivity titanium as a principal 
construction material. Titanium has some advantages in strength and cost over oxygen-free, high-purity copper, which has 
been widely used in many previous low-background experiments.  Radioactivity from photomultiplier tubes and materials 
used in construction of the TPC also needs to be reduced as much as possible.  This detector also needs a neutron 
shield, and a gadolinium-doped liquid scintillator is proposed to serve as a neutron veto detector.

Next-generation two-phase liquid argon (LAr) dark matter detectors have the same problems with purification and high 
voltage as the xenon dark matter detectors but they have the additional problem of reducing radioactive $^{39}$Ar. 
Pulse shape analysis can be used to identify $^{39}$Ar if rates and backgrounds are sufficiently low.  Argon from 
underground wells, which has much lower concentrations of $^{39}$Ar than atmospheric argon, is being used by the 
DarkSide family of experiments.  Detecting the 128 nm scintillation light from LAr requires wavelength shifting and 
efficient, large area photodetectors.  There are other very large (multi-ton) liquid noble gas detectors planned, 
which combine xenon and argon (XAX, MAX, Darwin)~\cite{XAX:2012}.  These experiments will have the same scale-up 
problems as the next-generation argon and xenon detectors.

The COUPP experiment plans to upgrade from 60 kg of refrigerant, running in SNOLab, to 500 kg with a 
possible move to Sanford Lab~\cite{COUPP500}.  It does not anticipate major scale-up problems other than keeping
the amount of radioactive contaminants in the detector as low as possible.

Another approach to WIMP detection is to look for a sidereal anisotropy signal. 
Most contemporary approaches, such as the DMTPC~\cite{DMTPC} and DRIFT~\cite{DRIFT} collaborations, attempt to 
visually track images in a low-density TPC. Both the DMTPC and DRIFT collaborations have built and operated track 
detectors for dark matter with  $\sim$0.1~kg or more of target material. With this small target mass, 
DRIFT was able to set limits comparable, at the time, to much larger non-directional detectors owing to these 
detectors' excellent discrimination ability. Both DRIFT and DMTPC are developing ideas to field detectors at 
the $\sim$1~kg scale, which would be capable of confirming or rejecting purported low-mass WIMP signals.  
An array of several hundred such detectors could provide clear confirmation to a dark matter detection 
by a generation-2 counting detector. Construction of such an array would be very challenging, but would be 
justified by the importance of providing an unambiguous signal for particle dark matter. 
R\&D on both electronic and optical readout will continue for the next several years, at which time the dark matter community 
will most likely gel around a common method for a large mass array.

R\&D on a columnar recombination approach may permit sensing of the nuclear recoil in a dense xenon gas and with 
a detector gas density $\sim$~100 times greater than possible with track visualization.  This idea has, 
at the moment, not been experimentally verified and the work is still in an early stage compared with the 
low-density, low-pressure TPCs described above.

\subsection{Light dark matter - axions}
\subsubsection{Current and planned projects}

Axions are another possible candidate for dark matter.  They are well motivated as a solution to the strong CP problem.
Axions are predicted to be light pseudoscalars with masses in the range of 10$^{-6}$ to 10$^{-2}$ eV, with the smaller 
mass preferred from astrophysical considerations, and a coupling to two photons.  There is currently one axion experimental search, 
ADMX, which uses the inverse Primakoff effect\footnote{$\gamma$ + axion $\rightarrow \gamma'$} in a tunable microwave 
cavity with an 8 Tesla magnetic field~\cite{vanBibber:2013ssa}. Since the energy of the outgoing gamma ray from an axion 
conversion includes the mass of the axion, the experiment searches for a resonant peak in frequency, using a radio receiver, 
as the cavity is tuned through possible axion masses.
 
\subsubsection{Key issues and next steps}

The major challenge in the ADMX experiment has been to develop a radio receiver with sufficient 
sensitivity to measure very 
low axion conversion rates in the microwave cavity.  The current ultra-low noise SQUID-based radio 
receiver is approaching 
the quantum limit for sensitivity.
A second major challenge has been to maintain a stable, high magnetic field as the microwave cavity is tuned over a 
large range of possible axion masses.  ADMX is well suited to detecting axions in the mass range 
10$^{-6}$ to 10$^{-5}$ eV, 
but larger axion masses are difficult to measure with this technique due to the difficulty in building 
suitable tunable cavities.

Fermilab is conducting R\&D on the use of a solid Xe detector operating at sub-Kelvin temperatures with superconducting sensors
to read out photon, ionization, and phonon signals.  This type of detector can be used to search for solar axions, which may be
created in the core region of the sun.  The axion-photon conversion by the strong Coulomb field of atoms in the crystalline Xe
detector may cause coherent Bragg scattering of the photons, which depends on the energy and incident angle of the axions.  This
strong correlation provides sensitivities below an axion mass of 10 eV, providing coverage of a QCD axion parameter space not
accessible to other experiments. 

\subsection{Cosmic microwave background radiation}
\subsubsection{Current and planned projects}

The Cosmic Microwave Background (CMB) is a critical component for understanding cosmology and exploring 
the fundamental physics of the early Universe~\cite{CMB:2013}. 
The focus of current CMB experiments is to characterize the faint CMB $B$-mode polarization pattern. 
There are two sources of $B$-mode polarization in the CMB. 
The first source, and the majority of the B-mode polarization, is due to 
gravitational lensing of the intrinsic CMB polarization, which appears at small angular scales 
($\sim$10~arcmins).  This lensing signal, which has recently been detected for the first time~\cite{CMB:Bmode}, 
provides a measurement of the growth of large-scale structure. 
Precision measurements of CMB lensing will provide constraints on the sum of the neutrino masses 
at a level relevant for understanding the neutrino mass hierarchy. 
The second, much harder to detect source of $B$-mode polarization is caused by the 
gravitational waves produced at the end of the inflationary epoch.  These perturbations produce an angular 
$B$-mode power spectrum that peaks at angular scales of $\sim$2~degrees. 
The amplitude of this signal is a measure of the energy scale of inflation, 
which is expected to be $\sim$10$^{16}$~GeV, the energy scale favored by Grand Unified Theories. 
The measurement of the inflationary $B$-mode signal would not only validate the inflationary paradigm, 
but would also constrain the physics of inflationary processes, physics not accessible by other means. 

CMB experiments are sensitivity-limited. The noise in an individual detector element is dominated by 
the shot noise of the 
measured photon flux. Experiments can increase sensitivity only by increasing the number of optical 
modes that are measured. 
The world-wide CMB program can be classified into ``stages" where each stage corresponds to a 10-fold 
increase in the number of optical modes and corresponding sensitivity.  
Stage-I experiments have been deployed and consist 
of focal planes that measure much less than 1000 optical modes.  
Stage-I experiments do not have the sensitivity to measure CMB $B$-modes. 
Stage-II corresponds to current operating experiments which measure on the order of 1000 optical modes. 
These operating state-of-the-art instruments have observed CMB $B$-modes for the first time. 
Bolometers based on Transition Edge Sensors (TES) dominate the stage-II technologies.   
Stage-III projects are in the R\&D stage and will deploy in the first half of 
this decade. These proposed experiments will measure on the order of 10,000 modes and will be the 
first to realize the era of ``precision CMB $B$-mode science.'' 
Technologies being pursued for stage-III experiments include arrays of single-moded TES 
bolometers coupled through superconducting microstrip and large-area multi-moded semiconductor bolometers. 
TES-based bolometers 
continue to be a core technology for stage-III.  A stage-IV experiment will measure on the order of 100,000
optical modes across a broad 
range of angular scales from a few degrees to a few arcmins and large fractions of sky ($\sim$20,000 deg$^2$, 
or about half of the full sky).  
TES-based bolometers with single-moded coupling are the only stage-III technology capable of 
simultaneously probing inflationary $B$-modes at large angular scales and lensing $B$-modes at 
smaller angular scales, which makes single-moded TES-based bolometers a critical 
technology for realizing stage-IV CMB science.

\subsubsection{Key issues and next steps}

The challenge for implementing single-moded TES-based bolometers for a stage-IV CMB experiment is one of scale. 
The key issues can be broken down into two classes: (1) stability of fabrication and 
(2) increased production throughput. Incremental and 
revolutionary R\&D will address both of these challenges, though the challenge of production throughput 
also requires expanded tools and fabrication resources along with high throughput test facilities. 
Incremental R\&D consists of stabilizing and scaling 
the fabrication of current detector architectures and materials, (such as detectors coupled to 
superconducting microstrips 
made from Nb-SiOx-Nb trilayers). More revolutionary development consists of pursuing novel materials, 
fabrication processes, 
and new detector architectures. In addition to detector R\&D, developments in the SQUID-based TES readout 
and multiplexing 
also have the potential to further simplify and streamline detector electronics.

\subsection{Tests of space-time}
\subsubsection{Current and planned projects}

An experiment at Fermilab, the Fermilab Holometer, will probe Planck-scale quantum geometry via position 
measurement with Planck spectral density.  Its non-local, bi-directional position measurement will probe non-commutative 
geometry~\cite{Hogan:2009}. 
The experiment correlates signals of nearly co-located 40-meter Michelson interferometers which can measure motions of 
$\sim$10$^{-18}$~m in $\sim$10$^{-6}$~sec.  When the ends of the arms of the interferometers are in a nested configuration,
space-time volumes will collapse into the same state and be correlated, while when the ends of the arms are in a back-to-back
non-overlapping configuration space-time volumes will be uncorrelated.

The holometer is currently under construction.  One of the keys to distinguishing ``signal noise'' from background noise is 
to look at the frequency spectrum of the noise.  The expectation is that signal noise will be frequency independent whereas 
mechanical background noise will fall off inversely with frequency, and the noise will not have the expected white noise 
spectrum.  To keep electronic and laser noise uncorrelated, independent lasers and electronics are 
used in each interferometer.

\section{Technologies}
\label{sec:instrum-tech}

The preceding three sections have demonstrated the range and breadth of the current and planned 
U.S. experimental HEP program.  
These experiments share the need for many technologies which are being developed or which are already in use.  
Accelerator experiments need:  
fast, highly pixelated, radiation-hardened,  low-mass vertex detectors with customized electronics;
low-mass trackers with fast, highly segmented readout and customized electronics;
cost-effective, highly segmented, radiation-hardened calorimeters with good energy resolution; and 
high-speed data acquisition systems.  
Experiments studying particle interactions with small cross sections or rare decays typically need 
cost-effective large 
volume detectors with affordable large-area charged particle detection or efficient optical readout systems.  
A subset of these experiments need materials and sensors with extremely low 
intrinsic radioactivity, while another subset need high-granularity detectors with fast timing and good 
energy resolution.  
Experiments in the Cosmic Frontier frequently need large arrays of ultra-low-noise electromagnetic 
radiation detectors.
The technologies described in the following sections have been identified as being generic to many 
HEP experiments.  They 
provide examples of areas where investment in detector R\&D could increase the physics reach, and/or reduce the cost 
of experiments in the current and planned HEP experimental program. A more detailed overview is given 
in a sensor compendium~\cite{sensors}. 

\subsection{Pixelated tracking}
Low-mass, radiation-hardened pixelated detectors will be needed for ATLAS, CMS, an ILC, CLIC and muon collider 
in the Energy Frontier, and LHCb and next generation B or $\tau$-factories in the Intensity Frontier.  
These will also often require sophisticated in-pixel processing such as time stamping, cluster finding, or 
inter-cell hit correlation. There are often competing demands for low-power, high-speed, 
electronics density and radiation hardness. 
The current technology of choice is bump-bonded hybrid pixels with either planar or 3-D silicon. 

\subsubsection{Monolithic pixels}
The performance of silicon-based pixelated detectors is defined by the pixel size, density 
and power consumption of the readout electronics, thickness of 
the sensor, and capacitance of the sensing nodes.  There are a number of technological approaches to 
the fabrication of detectors which can combine lower mass and power, decreased sensor thickness, 
and complex functionality for tracking detectors for HL-LHC and beyond.

Monolithic Active Pixel Sensors (MAPS) can evade sensor thickness and pixel size limitations by utilizing a thin 
(typically 15~$\mu$m) epitaxial layer under the CMOS transistor structures as the sensitive 
medium~\cite{MAPS:2006}.   
These devices can be made very thin, with small pixels, and avoid bump bonding.  
The main issues for MAPS sensors are radiation hardness and rate capability. 
Current MAPS sensors typically collect charge by diffusion rather than drift, making them about a factor of ten 
slower than biased sensors.  MAPS sensors used in commercial cameras have 5 transistors per pixel whereas HL-HLC
 rates will require  logic densities greater than 1000 transistors per pixel. 
They also suffer from parasitic charge collection from PMOS transistors fabricated 
as part of CMOS circuitry as well as the coupling of large digital signals from the CMOS 
readout circuit to the connected sensor region.

Current  MAPS R\&D has an emphasis on charge collection by 
drift instead of diffusion, which is mandatory for HL-LHC radiation levels.
High-resistivity epitaxial substrates can fully deplete and serve to mitigate 
radiation damage effects. HV-MAPS capacitively couples a MAPS-based 
sensor to a standard pixel readout chip, avoiding bump bonding and lowering cost~\cite{Barbero}. 
The T3-MAPS technology uses a buried n-layer to 
isolate transistors from the substrate and preserves high logic density~\cite{Mekkaoui}. 

Silicon-on-insulator (SOI) sensors utilize a thin (typically 200 nm) CMOS layer, oxide-bonded to a thick 
silicon handle wafer~\cite{Hara:2011hia, SOI:2008}.  
An SOI detector utilizes the handle wafer as a sensor connecting to the top electronics by vias etched 
through the thin 
oxide. These devices can be fully depleted, have fine pitch and low capacitance, and produce a large signal but 
are unlikely to be sufficiently radiation hard for HL-LHC.

Three-dimensional integration is a set of technologies for vertical interconnection of wafers including
bonding, thinning, and interconnection~\cite{Deptuch:2010zz}. 
It can offer advantages similar to those in SOI and MAPS applications without some 
of the disadvantages.  
For pixelated sensors it offers the opportunity 
for fine pitch (4~$\mu$m has been demonstrated) interconnection, low capacitance, and separate optimization of 
sensor, analog and digital layers. 
Silicon wafer stacking technology can be used to separate digital and analog circuits 
from the MAPS sensor, allowing 
for separation of PMOS and NMOS, eliminating parasitic collection, and allowing the sensors and analog circuits 
to be shielded from digital noise.
Principal issues for R\&D are related to the availability of the technology in 
foundries, and the issue of fabricating such devices with large area and low cost~\cite{Yarema:2008zza}. 
These developments are relatively recent and happening in several places at once.  The U.S. could establish a 
leading role with modest support for such activities.

\subsubsection{Radiation-hardened pixel sensors}
LHC pixel detectors need to include sensors capable of surviving radiation doses of about 
2 $\times$ 10$^{16}$ n$_{eq}$/cm$^2$ or providing easy 
replacement of the innermost pixel layers.\footnote{n$_{eq}$/cm$^2$ is 1~MeV neutron equivalent fluence.}
In silicon detectors the primary radiation effects are increased 
leakage current, introduction of p-type impurities, and charge trapping.  Some of the effects of p-type impurity 
introduction can be mitigated by device and defect engineering, but leakage current increase is a universal effect 
that can only be addressed by operating at temperatures of $\sim$ -20$^{\circ}$C.  For radiation doses of about 
$10^{15}$ n$_{eq}$/cm$^2$ the trapping 
distance will be significantly smaller than the thickness of a typical 300~$\mu$m thick detector.  One important 
method of mitigating radiation effects in silicon is to reduce the drift distance by reducing sensor thickness.  The 
depletion voltage is proportional to the square of the thickness, so the operating voltage is reduced and the reduced 
volume also reduces the leakage current. 3-D sensors take advantage of this by etching electrodes through the silicon. 
These electrodes then collect charge from the full detector thickness, while the small electrode spacing means that 
the devices can be fully depleted at low voltage~\cite{Parker:1997vy}.  

Diamond sensors are still being studied for certain applications where both radiation hardness 
and moderate-temperature operation are important. 
There are continuing concerns regarding material availability and decreased signal 
collection after irradiation due to trapped charges forming a polarization field~\cite{Trischuk:2009}.  
The smaller charge collected in diamond with respect to silicon requires the development 
of new, very low threshold readout chips. 

\subsubsection{High-speed sensors}
An ideal particle detector has low mass, large area, and includes pixelated tracking layers with both 
high resolution 
and high speed.  Time resolution scales as $\sigma_t = T_{rise} \times  \frac{Signal}{Noise}$, 
so a large-signal, low-noise system is needed.  
This can be accomplished by reducing the noise, by decreasing the sensor capacitance (which 
is not always possible), by increasing the amplifier current, or by increasing the signal.  In a silicon sensor, an
improved signal-to-noise can be obtained by increasing the detector thickness or by utilizing avalanche multiplication 
in the silicon itself~\cite{FastSilicon:2012}.  A sensor utilizing either proportional or Geiger 
amplification can provide very good time resolution at low analog power, but these devices typically suffer from high noise 
rates.  Solutions to noise might include cooling the sensors to reduce dark currents or utilizing a 3-D stack of SiPM-like 
sensors in coincidence to reject random noise.

\subsubsection{Large-area arrays}
Another important issue to resolve is how to build affordable, large-area arrays of pixelated detectors. 
Readout chips are typically limited to a reticule size of $2 \times 3$ cm$^2$ and by the overall yield, 
which scales with chip area.  Depleted silicon sensors typically require an inactive area larger than 2 to 3 times 
their thickness to reduce leakage currents near the diced edge.  
Active edge technology which utilizes 3-D sensors can eliminate this inactive 
area.  Undepleted MAPS sensors also do not suffer from this problem.  Technologies that combine 3-D sensors and 
interconnect technologies may provide low-cost ways of wafer-scale pixel tile production. Fine-pitch interposers of 
silicon or glass with through-vias can also provide a solution that matches the large areas available in wafer scale 
sensors and smaller areas associated with chips~\cite{Deptuch:2013soa}.   


Micro-pattern gas detectors (MPGDs) for charged particle tracking and muon detection are an alternative 
to pixelated silicon vertex and tracking detectors~\cite{Hohlmann:2013aga}.  
These  low-mass detectors have the potential of economically covering 
large areas and providing high tolerance against radiation damage, high spatial resolution 
(of order 10 $\mu$m), and good time resolution (of order 1ns).  
Future work is needed to reduce the readout cost by 
developing highly integrated, radiation-hardened front-end readout electronics with at 
least 4000 channels/chip and 
high-density detector electronics interconnections with flex-circuit readout electronics directly 
integrated into the MPGD structure. Work is also required to develop innovative signal 
induction structures to reduce cost, improve performance, and provide stability against electric breakdown. 
It is important to develop materials with resistance to aging and radiation damage, as well as cost-effective
MPGD construction techniques for high-volume production. 
There is a promising effort, led by the nuclear physics community, to establish an MPGD industry
base in the U.S. to produce large area foils. With the recent increase in activities by both the 
nuclear and high-energy physics communities, e.g., the planned ATLAS and CMS MPGD upgrades for the 
HL-LHC, this is a promising area where the U.S. could play a leading role
in the further development of this technology, synergistic between the two branches of science.


\subsection{Mechanics and power}
Mechanics and power refer to the interconnected issues of precision support structures, cooling, 
and electrical and thermal services.  The implications and issues have some commonality across the 
frontiers but also present some specific challenges.
The Energy and Cosmic Frontiers need very stable and precise structures, often with minimal mass
and high radiation length, to support tracking or imaging sensors. The Intensity and Cosmic Frontiers 
need large vessels, cryostats, and support frameworks, which may be possible with 
standard mechanical engineering approaches but may also require very radiopure materials in novel 
configurations.

Across the Frontiers, issues of placement, metrology, and alignment, both in fabrication and installation, 
are of considerable significance.  Precision mechanical actuator technologies can enable in-place 
alignment and adjustment of detector and sensor elements, such as tracker modules and fibers and imagers 
in focal planes.  Mechanical actuators also play a role in the assembly and placement process for sensors 
and modules.  The development of new types of probes, novel measurement configurations, and analytical methods 
are actively being pursued.

The goal of minimizing detector mass and maximizing radiation length requires an integrated system 
design incorporating cooling and power distribution as primary design considerations.  Radiation-hardened 
trackers may need to operate at temperatures as low as --30$^\circ$C.  For such applications, low-mass, low-Z, 
high-strength, thermally conductive materials and cooling structures are crucial.  A coordinated effort to 
develop new temperature-tolerant materials is key to science across multiple frontiers and has dramatic 
spin-off potential. 


Carbon composites are one example.  The U.S. currently has a dominant role in development and fabrication of 
carbon composites, with world-leading facilities and expertise at FNAL, BNL, and LBNL.  Additionally, 
thermally conductive carbon foam is a successful U.S. industry SBIR development --- it did 
not exist 10 years ago and is 
now commercially available. However, mechanical design and construction of composites 
require highly specialized engineering with a similar degree of specialization as IC design.  
Maintaining this engineering expertise is the only way to maintain leadership. 

Providing power and signals to future particle detectors, particularly front-end detectors in high radiation 
and high magnetic field environments, is an increasingly challenging cabling problem since thousands 
of readout chips require tens of thousands of amperes of current at low voltage and 
large multidrop or point-to-point 
signal distribution. This requires appropriate cable and insulation materials, including large
area flex cables, micro-coaxial cables, low-mass conductors, low mass reliable connectors, and encapsulants.

The main approaches to power distribution include DC-to-DC conversion and serial current flow.   
DC-to-DC converters at the load point allow power to be delivered at higher voltage 
and lower current. Radiation hardened DC-to-DC converters with air core inductors can operate in high 
magnetic fields but are currently limited in step-down ratio and efficiency. For example, the capacity 
of the CERN DC-DC converter is the central limiting factor in the design of the CMS Phase-II tracker 
readout chip. There is potential for very significant improvements in voltage ratio and efficiency 
as well as radiation hardness by utilizing laterally diffused CMOS converters, 
utilizing gallium nitride transistors, 
and moving to more efficient designs operating at higher frequency~\cite{Dhawanpower}.  
Serial powering works best when 
multiple loads all demand the same current.  Novel configurations that optimize the current 
distribution are an interesting area for further study.  Circuits and systems to monitor, control,
and protect the power distribution network are also critical.  Point-of-use circuits to reliably 
multiplex and monitor the high voltage bias current delivery to sensors can dramatically reduce 
the number of cable plants servicing large detector arrays and reduce the number of power supplies.

The development of radiopure materials covers both basic materials such as copper, titanium, and 
semiconductors and the fabrication of novel and specialized components.  For example, the emerging 3-D 
printing technologies, combined with new radiopure polymers and ceramic or metallic powders, could be 
very relevant to future low-background applications~\cite{Hohlmann:3D}.  A community-wide or small 
business approach to radiopure materials, fabrication, and assay could be very beneficial in this regard.

\subsection{ASICs}
Application Specific Integrated Circuits (ASICs) were first developed for particle physics in the 
1980s to read out silicon strip vertex detectors in colliding beam experiments. 
They have since become one of the core technologies available to detector designers.
A separate Snowmass study has been carried out to evaluate integrated circuit design in the U.S. 
and its findings and recommendations are reported in the white paper 
``Integrated Circuit Design in the U.S.''~\cite{DeGeronimo:2013zxa}. A summary of the key points
is provided below.
 
A number of factors make ASICs essential to HEP. These include:
\begin{itemize}
\vspace{-4mm}
\itemsep0em
\item \textbf{Small physical size}: The space constraints of many detectors, most notably pixel vertex detectors, 
 require custom microelectronics. Even when commercial electronics can be used, small ASICs 
can often be positioned closer to the sensor than would otherwise be possible. This reduces input 
capacitance and improves noise performance. In many cases the size of the cable plant is also 
reduced (e.g., smaller radiation length, less power dissipation on the detector, lower cooling).
\item  \textbf{Low power dissipation}: The infrastructure required to power and cool on-detector electronics 
often limits detector performance. Especially in high channel count applications, low power dissipation 
can be one of the most important specifications.
\item  \textbf{Radiation tolerance}: Many HEP applications require 10 - 100 Mrad total ionization dose  
tolerance as well as immunity against single event upsets. Future vertex detectors may 
require Giga-rad tolerance.
\end{itemize}
ASIC-related R\&D is required in a number of areas in order to improve science output or simply 
to make future experiments in the Intensity, Cosmic, and Energy Frontiers possible. 
Examples are~\cite{Yarema:2008zza, Deptuch:2013soa}:
\begin{itemize}
\vspace{-4mm}
\itemsep0em
\item high-speed waveform sampling
\item picosecond timing
\item low-noise high-dynamic-range amplification and pulse shaping
\item digitization and digital data processing
\item high-rate radiation-tolerant data transmission
\item low-temperature operation
\item extreme radiation tolerance
\item low radioactivity processes for ASICs
\item low power
\item 2.5-D and 3-D assemblies
\end{itemize}

Major recommendations of the report include the continuation of strong physicist--engineer links, 
an increase in generic ASIC R\&D, education to provide basic literacy in IC technology for physics 
students, and increased communication and collaboration among U.S. laboratories and universities. 
The DOE might want to consider an ASIC stewardship role for HEP, analogous to the particle 
accelerator stewardship role.

\subsection{Calorimetry}
Electromagnetic and hadronic calorimetry is central to many experiments in all Frontiers. At the LHC the 
principal challenge to detector subsystems is the survivability in the radiation field and at high rates. 
In addition, the ability to detect and 
accurately measure electromagnetic showers in the presence of large backgrounds is a major challenge
faced in calorimeter designs for the HL-LHC. 
Intensity Frontier experiments utilize  calorimeters designed 
to detect and measure electromagnetic showers with good resolution at energies from 10 GeV  down to 
several MeV~\cite{pxemcal}. 
These devices often need precision timing in the range of a few tens of picoseconds, 
while having a high tolerance to ionizing radiation. 
Future lepton collider calorimeters are required to have excellent energy resolution 
both for single particles and for jets. Size, cost and containment are critical. 
For applications in space, calorimeters need to be compact, omni-directional, 
and able to distinguish between electrons, gamma rays, hadrons, and high-Z nuclei. 

In the last two decades  there have been revolutionary changes in detector technology, which 
significantly impact calorimetry. The field has seen  
a significant increase in the capabilities of front-end electronics, enabling practical 
systems with a very large number of channels. 
There has been the development of very dense inorganic scintillators with large scale production 
capabilities, and enormous progress in understanding radiation damage and its mitigation. 
In the area of readout, robust, compact photodetectors such as silicon photomultipliers (SiPMs)
have been developed, 
capable of working in high magnetic fields and allowing the construction of hermetic, 
yet highly segmented calorimeters.
There has been continued progress in the development of modeling tools, allowing for detailed analysis of  
physics processes in the body of the calorimeter and 
advances in nanotechnology and non-linear optics allowing  for major improvements in 
light collection and detection schemes. 
These new capabilities have spurred development of promising techniques for hadron calorimetry, such as particle flow and dual readout.

These changes have led to a re-evaluation of our
approach to calorimetry.  The particle 
flow technique, pioneered by the CALICE collaboration,
is based on utilizing the momentum resolution provided by 
the inner tracking system to provide the basis for precise jet reconstruction. Information from the tracker
can be combined with a fine-grained calorimeter to identify shower components with 
charged tracks, locate energy deposit from neutral hadrons, and combine this information 
into the best estimate of jet energy.  Simulations have been shown to reconstruct jets at the
$Z$ boson mass with a precision of 3\%.  While this approach was developed
in the context of experiments for ILC and CLIC, recently the
adaptation of these ideas to the domain of hadron and muon colliders
has begun.

A separate development, which has also resulted from a better understanding
of the physical processes of hadronic showers, is the concept of dual
readout calorimetry. In this approach, pioneered by the DREAM
collaboration~\cite{DualReadout:2013}, correlations between the electromagnetic and hadronic fraction of
hadronic showers are utilized to correct for energy lost to nuclear breakup. 
This is done by making separate measurements of Cherenkov and scintillation 
light in a shower. Thus, by knowing on an event-by-event basis how
much energy has been deposited electromagnetically, fluctuations in the
hadronic component can be corrected to achieve much better energy resolution 
than provided by the separate Cherenkov and scintillation measurements.

At the LHC, the CMS detector demonstrated the possibility of operating a
crystal calorimeter at a hadron collider even with the lattice damage that
occurs with hadron interactions, while the results from the ATLAS liquid
argon calorimeters have shown that sampling in depth in a shower helps
significantly in reconstructing the physics events.

There are other R\&D efforts related to calorimetry technology.  
Secondary emission detectors and novel radiation-resistant
wavelength-shifting or scintillating fibers are two active 
areas of R\&D~\cite{Albayrak-Yetkin:2013xga}\cite{Albayrak-Yetkin:2013yma}. 

Brighter, faster and radiation hardened crystals, e.g., cerium doped lutetium oxyorthosilicate 
(Lu$_2$SiO$_5$ or LSO) and lutetium-yttrium oxyorthosilicate (Lu$_{2(1-x)}$Y$_{2x}$SiO$_5$ or LYSO), 
are needed for future HEP experiments at the Energy and Intensity Frontiers. 
With high density (7.1 g/cm$^3$), fast decay time (40 ns) and superb radiation hardness, 
LSO/LYSO crystals are the baseline for the Mu2e experiment at the Intensity Frontier and 
are being proposed for the CMS forward calorimeter upgrade for the HL-LHC~\cite{zhucal}. 

Fast crystals with sub-nanosecond decay time, e.g., barium fluoride (BaF$_2$) and YAP:Yb, 
are needed for future HEP experiments at the Intensity Frontier to cope with the unprecedented 
event rate. To facilitate $\pi^0$ reconstruction, a longitudinally segmented crystal electromagnetic 
calorimeter (ECAL) 
would provide a photon pointing resolution in addition to excellent energy and position resolutions. 

The cost of the crystals is a crucial issue at future lepton colliders because of the huge crystal
volume needed to build calorimeters.  UV-transparent crystals, such as PbF$_2$, PbFCl and BSO,
can achieve unprecedented jet mass resolution and are being considered for dual readout of
Cherenkov and scintillation light in a homogeneous hadron collider calorimeter (HHCAL)
detector concept.

R\&D for future crystal calorimeters thus will proceed along the following lines.
\vspace{-4mm}
\itemsep0em
\begin{itemize}
\item
Investigation on the radiation damage effects induced by gamma-rays, neutrons, and charged 
hadrons in various crystals to be used in future HEP experiments. 
\item
Development of super-fast crystals with sub-nanosecond decay time.
\item
Development of dense, UV-transparent, and cost-effective crystals for the HHCAL detector concept. 
\item
R\&D on detector design and compact readout devices for a segmented crystal calorimeter with 
photon pointing resolution.
\end{itemize}

\subsection{Photodetectors}

Photodetectors provide essential functionality for HEP experiments
in a wide variety of contexts. Light can be generated by particle
cascades in calorimeters, by scintillation in liquid noble gas
detectors, in gas avalanches, by Cherenkov radiation, by
scintillation or Cherenkov light in crystals, in extensive air
showers, or through scintillating plastics. Requirements vary, but
the physics reach of experiments in all three Frontiers can be
extended through targeted improvements in three key areas: (1)
spatial and time resolution, (2) economical coverage of large areas, 
and (3) robustness under extreme conditions such as high radiation
environments, cryogenics, and high pressures.

Many measurements in particle physics are limited by the time
resolution of particle detectors. Conventional photomultiplier
tubes are single-pixel detectors with time resolutions of order
nanoseconds. Next-generation photodetectors could deliver
sub-nanosecond time resolutions and finely segmented spatial
resolutions~\cite{Ossy_highres}--\cite{RonzhinTOF}.
 
Often, the combination of precision timing and spatial resolution
is as important as the timing itself, allowing the measurement of
event kinematics in terms of 4 vectors rather than 3 vectors.
Measurement of the arrival times of particles with resolutions
approaching single picoseconds over a drift distance of 1.5~m
could be combined with track information to determine particle
masses, and thus the quark composition, of hadronic jets. Timing
measurements can also be used to match high energy gamma rays to their
respective interaction vertices, providing a useful capability for
collider experiments interested in measuring certain Higgs, Kaluza-Klein,
and gauge-mediated SUSY signatures. Similarly, in neutrino
experiments where Cherenkov light is present, the combination of
precision position and time information can be used to reconstruct
tracks and EM showers with high resolution, which is important in
reducing backgrounds for electron neutrino selection, as well as
discriminating between gamma rays and electrons~\cite{{LAPPDneut},{LAPPDneutWP}}.

Heavy flavor physics experiments rely on technologies such as
Ring Imaging CHerenkov (RICH) and Direct Internally Reflected
Cherenkov (DIRC) detectors, where the geometric and timing properties of
Cherenkov light emitted by high energy particles passing through
thin radiators are used to separate particles of similar
momentum but differing mass. DIRC detectors, in particular, rely
on timing to correct for chromatic dispersion effects on these
Cherenkov rings~\cite{VavraTOF, VavraFDIRC, BellII-TOF, VavraPID}.

Given the large scales of future collider and
neutrino experiments, a substantial
improvement in the cost-per-area coverage of photodetectors can
have a positive impact on large experiments. While this is not the
only factor, improvements that allow better light collection and
area coverage per unit cost can address key physics needs within
the currently tight budgetary environment. Further gains can be
made using light collectors, devices such as Winston cones and
wavelength-shifting plastics. 
These allow for improved light collection with the same number of photosensors. 
Finally, in the context of large water Cherenkov experiments, as 
in flavor physics, it is possible to employ full imaging optics such as
parabolic and spherical reflective geometries.

A particularly fruitful area for future R\&D is the development of
novel, high-quantum-efficiency photocathodes, capitalizing on
advances in the field of materials science~\cite{Xie}.
It may not only be
possible to significantly increase the detection efficiency of
single photons, but also to tune photocathode responses for custom
spectral ranges~\cite{photocathodeworkshop}.

Many HEP experiments subject photosensors to extreme conditions,
particularly low temperatures, high pressures, and high-radiation
environments. Continued work on implosion mechanisms,
stress-testing, and containment are important for vacuum-photosensors 
in deep-water and ice applications. It is also
important that photosensors can survive and operate under
exposure to cryogenic temperatures, such as those of liquid argon.
In most collider-based environments, photosensors will be subject
to unprecedented radiation doses. These conditions require 
development of radiation-hardened materials, such as fused silica and optical
glues for DIRC-based detectors, SiPMT-based detectors with a good
resistance to large neutron doses~\cite{LehmanMCPs}, 
photocathodes with resistance to radiation, 
and many other materials.

The Large Area Picosecond Photodetector (LAPPD) collaboration has
developed flat-panel, large-area ($20 \times$ cm$^2$) micro-channel plate (MCP) 
detectors with
photocathode quantum efficiencies of $20\%$ or better, a gain of
between $10^6$ and $10^7$, $xy$ position resolution of about one 
millimeter, and time resolutions approaching single picoseconds.
These detectors consist of MCP-based detectors, 
made by thin-layer deposition of enhanced secondary electron emissive 
materials on substrates made from drawn glass capillary tubes. The vacuum 
package is based on flat-panel, sealed-glass technology, with no external 
pins. The delay-line anode pattern consists of silver micro-striplines,
where differential timing is used to measure the hit position in
the direction parallel to these striplines and charge centroiding
is used to determine the hit position in the transverse dimension.
The nominal LAPPD design addresses the entire detector system,
including front-end electronics. The LAPPD collaboration has
developed a low-power ASIC, capable of sampling fast pulse shapes
with 1.5 GHz analog bandwidth and up to 15~Gsamples/sec sampling
rates~\cite{LAPPDWP}.

Another area is the development of novel phototube
geometries and electron optics. One example is the Microchannel Plate-based 
Large Area Photodetector (MLAP) project in China, which uses a 
perfectly spherical geometry, along with both
reflective and transmissive photocathodes to increase photon
detection efficiency~\cite{MLAP}. 

Multi-gap resistive plate chamber (MRPC) technology, as pioneered by
the ALICE detector, may be useful for very large time-of-flight (TOF) systems.
Its TOF resolution is close to $\sigma \approx$ 50~ps/track, which is a 
great achievement, considering that the detector's total area is close to 
$\approx$~150~m$^2$.  It provides a good particle-identification (PID) performance 
below ~1 GeV/c. With more gaps in the MRPC multi-electrode structure, timing 
resolution could move towards $\sigma \approx$~10-20 ps/track. The STAR 
experiment reached a timing resolution of better than 100~ps/track.  Its total 
area was ~50 square meters~\cite{{TOFMRPC},{STARMRPC},{newMRPC}}.

\subsection{Data flow}
Acquisition of massive amounts of data is a problem in all physics frontiers.  However, 
in spite of significant differences in operating regimes, the data acquisition (DAQ) demands from different 
frontiers have similarities that can be generalized to the need of processing many data 
links at rates of between 3 and 10 Gigabits/second.

Luminosity upgrades to the LHC present significant challenges to existing LHC experiments. 
Operating the HL-LHC with 25 ns bunch crossing spacing at $5\times10^{34}$cm$^{-2}$s$^{-1}$ 
implies a pileup of up to 200 minimum-bias events per crossing. This is an order of magnitude 
greater than the 17 minimum-bias events per crossing at the LHC design luminosity ($10^{34}$cm$^{-2}$s$^{-1}$), 
and will degrade all occupancy-dependent trigger algorithms that rely on forms of 
isolation to identify electrons, muons, taus, and missing-energy signals.  
This requires a better-performing trigger with additional information, such as tracking data, 
which is used to reduce the trigger rates against the much higher backgrounds. The size of regions 
sampled for trigger decisions will need to shrink to handle the increased backgrounds. Therefore, 
substantial improvements will be needed for the ATLAS and CMS trigger systems for the HL-LHC.  
The ATLAS and CMS detectors will acquire 2 MB of data every 25 ns, or 10 TB/s.  
This requires a sophisticated trigger system, since this amount of data cannot be read out.  
ATLAS and CMS anticipate a Level-1 trigger system that reduces the input data rate to between 
0.2 - 1 MHz for readout into the higher-level triggers, which output their data at 5 - 10 kHz.  
These experiments require studies of high-speed links, field programmable gate arrays (FPGAs), 
and high-speed backplane infrastructures in order to meet these challenging requirements. 

The LSST $\sim$3 billion pixel camera needs only a 1 Hz readout rate, but each readout consists of 
gigabytes of data.  An increased bandwidth for the LArTPC readout for LBNE can allow a triggerless 
readout with minimal front-end sparsification to avoid unnecessary loss of precision and can defer 
processing to sophisticated algorithms running on online or offline computers.  
Modular component high speed massively parallel DAQ systems are one possible solution. 

A promising modulator-based,\footnote{Mach-Zehnder and absorption modulators with 
forgiving stability requirements.} high-bandwidth, optical readout in-air 
system for HEP detectors is in development at ANL.  Optical modulators are more reliable, 
are lower-cost/Gigabit/second, are more radiation tolerant, 
and operate at lower power than vertical-cavity surface-emitting lasers (VCSELS). There are also 
small business development projects to develop radiation hardened, low-mass modulators.
In addition, studies with wireless RF readout and optical beam power transmission have been carried out 
to test the application of new wireless readout technologies for use with large detectors or with 
smaller detectors that require increased mobility, such as detectors deployed to 
monitor the reactor neutrino flux or measure a neutron flux in security applications. 
The front-end circuitry, including a high-voltage power supply, is powered wirelessly, thus 
creating an all-wireless detector system~\cite{wireless}. 

Chips that can perform waveform sampling at rates of gigasamples/second are also useful in DAQ 
by allowing experimenters to perform real-time or offline digital signal processing, pulse shape analysis, 
cross-talk correction, and precision timing extraction.  While commercial waveform digitizers can often provide 
this functionality, custom waveform sampling ASICs have been developed for applications requiring some combination of
multi-gigasample/second sampling rates, low-power, and very high-channel density.  However, these custom chips need 
very large buffer depths to handle fast sample rates and relatively slow readout rates, which can be a problem.  In
addition, many timing calibration constants for the large numbers of delay lines are needed for the fast 
sampling rates. 

\subsection{Large-volume, fully sensitive detectors}
Large-volume, fully sensitive detectors are crucial for both neutrino and dark matter studies.  The focus for 
neutrino physics is liquid argon, but other materials such as xenon, bulk semiconductors, or high-pressure gases 
are being studied.  Dark matter experiments utilize semiconductors as well as liquid or solid noble gases or 
crystals. 

Future high-mass neutrino detectors are likely to be fully sensitive noble liquid or water-based detectors.  
In order to fully exploit noble liquid detectors, we need to improve our understanding of the basic physics 
of the processes involved in producing the observed charge and light signals, as well as the backgrounds 
that accompany the desired signals. We also need a dedicated program to establish the properties of liquid argon, 
such as diffusion, electron-ion recombination, and electron attachment to impurities, as they relate to the desired 
physics capabilities of the detector and studies of other issues related to TPC construction, such 
as optimal wire spacing, the use of cold electronics, the reduction of backgrounds from contaminants, and 
shielding from cosmic rays.

Dark matter experiments are ultra-low in rate, less than $10^{-3}$/day/kg, and very low in energy, between 
10 -- 100~keV, making background reduction and rejection the overriding experimental consideration. 
Most experiments utilize dual signals -- light and ionization for noble liquid detectors, or phonons 
and ionization for semiconductor detectors -- to separate the dark matter signal from radioactive backgrounds.  
Cosmic Frontier experiments span a 
very large spectrum of technologies and techniques, but they often share the requirement for low and well known 
background levels.  A HEP community-wide or small business approach to radiopure materials and assay is 
likely to be very beneficial to a variety of experiments.  Many dark matter experiments are planning to scale up in 
area to improve sensitivity.  If a signal is found, the next step may be a much larger-scale dark matter observatory. 
An important capability for any such detector would be the ability to identify the direction of the nuclear recoil 
and therefore the WIMP wind. One possible approach is to utilize the ratio of recombination and ionization rates 
in the dense ion trail surrounding a recoil nucleus as a function of external field direction.

\subsection{Low background materials}

Direct-detection dark matter searches and neutrinoless double-beta decay experiments require very low radiation 
backgrounds and must be constructed with materials that are extremely radiopure. Controlling background 
radioactivity is one of the major requirements of the next generation of these experiments~\cite{lbkd}.  
In general, this 
is accomplished by (1) selecting materials that are known by their chemistry to have intrinsically low levels 
of radioactivity and are free from contaminations of elements such as U and Th~\footnote{U and Th produce neutrons 
via spontaneous fission or ($\alpha$,n), depending on the matrix in which they are embedded, as well as gamma rays 
with energies up 2.614 MeV.}, (2)~paying special attention to handling these materials and reducing their exposure 
to sources of radioactive activation and surface contamination, and (3) assaying the materials for radioactive 
isotopes by gamma counting with high-purity germanium detectors or testing small samples using mass spectroscopy.  
In addition, modern metallurgical techniques and special attention to polymerization precursors can reduce 
radioactive contamination in materials.  For example, high-Z construction materials can be purified using 
electrochemical techniques to levels resulting in $<$ 1 $\mu$Bq/kg $^{238}$U and $^{232}$Th activities. 


\subsection{Superconducting sensors}
Our capabilities with superconducting technologies have become sufficiently mature that superconducting 
detectors can be routinely fabricated with nearly quantum-limited performance.  The commercial availability 
of cryogen-free, sub-Kelvin systems has significantly lowered the operational cost and complexity of 
operating superconducting technology.  Thus, it is possible to envision superconducting detectors having 
profound impact on future HEP science.  The principal challenge facing increased incorporation of 
superconducting detectors into HEP is one of scale.  Developing the capability of high volume production
and operating a broad spectrum of superconducting detector arrays is an important goal for HEP instrumentation.

There are two superconducting technologies that have the most promise as detector arrays.  The first of these
is Transition Edge Sensors (TES) technology.  Over the past 20 years, our understanding of TES detectors 
has grown substantially. TES detectors have led the past decade in dark matter direct detection and they 
have transformed the landscape of Cosmic Microwave Background (CMB) measurements. 
For both of these applications, the detectors have reached their fundamental background 
limit, and further improvements in scientific reach are possible only with larger detector arrays.  The 
frontier of research for TES technology is mass production of cutting-edge detectors along with increasing 
the density of their cryogenic readout.  

A second technology, kinetic inductance detectors (KIDs), has emerged at the turn of the century.  
Progress with KIDs has been rapid, and a detailed understanding of their micro-physics is expected 
within a few years. KIDs are a particularly attractive technology because they are naturally multiplexed, with 
a single element for both the detector and readout.  This attribute reduces the number of 
steps required in fabrication, enabling high yield and rapid production of detector arrays.

Developing the techniques for fabricating large arrays of superconducting detectors promises immediate impact 
at the Cosmic Frontier.  Making dark matter detector arrays with 100 to 1000 detectors will enable searches 
for low-mass WIMPs, along with ton-scale heavy-WIMP searches using superconducting technology. 
Making CMB bolometer arrays with 10,000 to 500,000 detectors will enable the next generation of CMB experiments, 
sustaining U.S. leadership in the field of CMB science. These experiments will explore the connection between 
inflation and Planck-scale physics while also measuring the sum of the neutrino masses.  Making this many 
optical imaging spectrometers will be revolutionary for future studies of dark energy by enabling simultaneous 
large-aperture imaging and spectroscopy.  Investment by HEP into generic superconducting instrumentation is 
critical for the development of these array technologies for HEP science.  The immediate scientific impact of 
this R\&D is at the Cosmic Frontier, though cultivating the expertise and resources for fabricating and reading 
out multiple kinds of superconducting detector arrays will likely have much broader impact. For example, there 
are applications of highly-multiplexed TES arrays for neutrino physics via measurement of coherent neutrino-nuclear 
scattering and detection of the cosmic neutrino background.

The majority of current efforts for developing superconducting detector systems is project-driven.  The 
closest semblance of a broad superconducting detector program is R\&D for space-based applications. Globally, 
there is no effort focused on developing generic superconducting technology for HEP science. The U.S. HEP 
community is best positioned to lead this development because of its connection to multi-disciplinary labs 
with the resources and tooling required for fabricating and reading out superconducting devices.  Developing the 
capability and expertise for mass production and operation of superconducting detectors promises quick 
scientific return at the Cosmic Frontier while also providing access to new technology to the broader HEP 
community and beyond. 

\subsection{Emerging technologies}
\label{sec:instrum-emerging}

There are a variety of emerging technologies and ideas for using conventional technologies to solve current or 
anticipated problems in HEP.  Several of these are based on novel materials, while others are based on using 
existing technologies in novel ways.

Graphene, a pure carbon planar structure, one atom thick with atoms arranged in a hexagonal structure, is of 
interest to HEP because of its extremely high electron mobility at room temperature, an optical opacity which 
is electrically tunable at frequencies up to a terahertz, very high thermal conductivity, very high strength 
and rigidity, and very low weight.  Potential applications include use in ICs, where its high 
carrier mobility and low noise make it an excellent material for channels in field effect transistors, and in 
field effect transistors on undoped substrates used as high resolution particle detectors.  Graphene's high 
electrical conductivity and electrically controllable optical transparency make it an excellent candidate for 
switching optical devices, such as modulators for optical data acquisition systems.  Considerable R\&D is 
necessary to produce inexpensive, usable quantities of graphene and devices based on it.  Silicene, a silicon 
analog of graphene, is also being studied for possible similar applications and has the advantage of being an 
intrinsic semiconductor for electrical applications.

Another material that may be useful for HEP particle detectors is amorphous nanocrystalline thin-film silicon. 
This material provides the advantages of amorphous silicon, such as its strong absorption in the visible light 
spectrum, radiation hardness, and ease of uniform deposition over large areas on a variety of substrates, while 
benefitting from low-cost thin-film fabrication technology.  To achieve desirable detector characteristics, such 
as long carrier lifetime and substantial drift velocity, nanoparticles of crystalline silicon are incorporated 
into the amorphous silicon, marrying the detector-grade characteristics of crystalline silicon with the inexpensive 
production methods of large scale amorphous silicon.  Preliminary results using hydrogenated amorphous silicon 
containing silicon nanocrystallite inclusions indicate large enhancements of mobility and recombination lifetime 
compared to hydrogenated amorphous silicon films.

The plasma display panel (PDP) used in plasma TVs is an existing technology which is being studied for possible 
use as a large, inherently digital, high gain variant of a micropattern gas detector.  The plasma panel sensor (PPS), 
which will be derived from the PDP, is comprised of a dense array of small, plasma discharge gas cells within a 
hermetically sealed glass panel, and is assembled from non-reactive, intrinsically radiation-hard materials such 
as glass substrates, metal electrodes, and mostly inert gas mixtures.  The PPS offers the potential to provide 
low-cost, large-area ionizing radiation detectors with high granularity (1~millimeter), good spatial resolution 
(300~$\mu$m), and fast response time ($\sim$10 ns).  Technology for fabricating these devices with low mass 
and small thickness, using gas gaps of at least a few hundred micrometers, is being developed.

 

\section{Facilities}
\label{sec:instrum-facilities}

\subsection{Test beams}

At present, there are two test beam facilities in the United States, one located at Fermilab,
and the other at SLAC. Each is heavily utilized for detector development.
  
The Fermilab test beam can provide 120 GeV protons, 32-66 GeV pions, 1-32 GeV mixed beam, and, 
when using a steel absorber target, a broadband muon beam.  It can provide fluxes from 100 - 5,000 Hz 
below 8~GeV and up to 100,000 Hz for energies up to 120 GeV.  It is instrumented with two Cherenkov 
detectors, a fixed beam telescope with a $\sim$10 micron tracking resolution, four multiwire 
proportional chambers with $\sim$500~micron resolution, two time-of-flight systems --- one with 
24~psec and one with 160~psec resolution --- and lead glass calorimeters and scintillators.  

SLAC has several test beams, one of which is appropriate for HEP detector R\&D.
It is an electron beam with energies between 2.2 and 15.0 GeV, a repetition rate of 120 Hz, and
a beam charge between 20 - 250 pC with a typical value of 150 pC.  There is a clean secondary
beam of electrons which can be tuned to energies between 2 and 13 GeV with 1 - $10^9$ e/pulse.
Two scintillation counters are used as a trigger, and there are three silicon strip layers and 
lead glass blocks for beam instrumentation.  

\subsection{Irradiation capabilities}

Irradiation facilities needed to test detectors up to radiation levels of $10^{16}$ 
particles/cm$^2$/sec consist of (1) radioactive sources that typically provide low energy $\gamma$ 
or neutron irradiation, (2) accelerated beams that can provide energetic protons, electrons and 
neutrons, and (3) nuclear reactors that typically provide low energy neutrons.  There is also a 
High-Intensity Gamma-ray Source (HIGS) facility at Duke University, which is a free-electron 
laser (FEL) based Compton backscattering gamma ray source.

Radioactive source irradiation facilities exist at Argonne and Brookhaven National Laboratories as well as a few 
universities.  In addition, Fermilab has several radioactive sources, including a 200 Cu 
$^{137}$Cs source which generates $\sim 3 \times 10^6$ gammas/sec or 700 rads/hour at a 
distance of 30~cm.  Sandia National Laboratory provides 
both gamma and neutron irradiation facilities. 
The Gamma Irradiation Facility (GIF) provides both $^{60}$Co and $^{138}$Cs irradiation; 
the Annular Core Research Reactor (ACRR) provides neutron irradiation~\cite{seidel}. 

Accelerated beam irradiation facilities include the 88 inch cyclotron for 
protons or ions at Lawrence Berkeley National Laboratory, a proton cyclotron at 
Massachusetts General Hospital, and  the Los Alamos Neutron Science Center (LANSCE) 800 MeV proton beam.
In addition there are the MT3 120 GeV proton, the MTA 400 MeV 
proton, and the 10~MeV neutron facilities at Fermilab, the Loma Linda University proton therapy 
machine, the Indiana University ISAT 30-200 Proton Cyclotron with fluxes from $10^2$ 
to $10^{11}$ protons/cm$^2$/sec, and the University of Massachusetts at Lowell 
irradiation facility which can produce between 100~rad to 2~Mrad/hour.

The LANSCE facility 
has a typical beam current of 83 nA or
$5 \times 10^{11}$ protons/pulse at a repetition rate of 1~Hz.  A high-intensity pulse mode
of up to 1 $\mu$A protons/pulse is also available with beam currents up to a microamp.  The
beam spot is quasi-Gaussian with diameters ranging from 1 to 8~cm.  

\subsection{Underground facilities}

In the United States, there are three principal underground facilities, the 4,850 foot deep (1,478 
meters) laboratory at the Sanford Underground Research Facility (SURF) in Lead, South Dakota; the Soudan 
Underground Laboratory at a depth of 710 meters in northern Minnesota; and the Waste Isolation 
Pilot Plant (WIPP) Underground Laboratory located in a low radioactivity salt deposit at a depth 
of 2,150 feet (655 meters) in Carlsbad, New Mexico.  There is also an underground research facility
at a depth of 1,700 feet (518 meters) at the Kimballton Underground Research Facility (KURF) in Virginia. 
We refer to Capabilities in Chapter 7 for a full overview of the world-wide facilities.   


\section{Universities and partnerships}
\label{sec:instrum-universities}

\subsection{University resources}

Universities provide crucial expertise, infrastructure, and resources to the particle physics instrumentation 
program.  They provide both depth and breadth of resources that cannot be supported at the national 
laboratories.  Universities have unique connections to scientists at the forefront of materials science, 
nanotechnology, chemistry, and electrical engineering, any of which might provide vital insight into the 
next breakthrough technology. They also provide training for the next generation of scientists and instrument 
developers and provide the graduate students and postdoctoral fellows who do much of the work of developing 
and testing new technologies.

Despite continuing erosion of university R\&D funding, there are still a large number of university-based centers 
of excellence that provide focused support for projects and R\&D in particle physics.  Some universities have 
machine shops capable of precise fabrication as well as fabrication of large objects, centers for design of 
integrated circuits, laboratories with expertise in bump bonding and circuit interconnection, laboratories for 
crystal development, and capabilities for silicon detector testing and radiation damage studies. University-based 
scientists and engineers develop integrated circuits for ATLAS and CMS, design and develop carbon fiber structures 
for pixel detectors at collider experiments, develop superconducting sensor technologies, provide the expertise 
to produce pixel detectors in U.S. university laboratories, and develop and test a variety of prototype detectors 
for use in large and small HEP experiments.

Partnerships with university nanotechnology and materials science departments have yielded radiation-hard 3-D 
silicon sensors now utilized in the ATLAS Insertable B Layer (IBL), techniques for thinning and annealing 
sensors as thin as 50~$\mu$m, transition edge sensors used in CDMS, and carbon fiber support structures used 
in the D0 and ATLAS detectors. 

\subsection{Partnerships}
\subsubsection{Research collaborations between industry, universities and laboratories}

National laboratories and universities are key drivers for research and innovation by using their facilities
and expert personnel to carry out fundamental and applied studies to advance basic knowledge.  
Industry growth is fueled by constant innovation to improve the performance and features of established 
products, as well as investing in disruptive technologies that ultimately create new products which can be 
cheaper, simpler, and/or smaller.  Collaborations between national laboratories and universities, 
which develop the disruptive technologies, and industry, which can invest in these technologies to produce 
better products, can be mutually beneficial.  

Because of the increased complexity of modern research experiments and the applications of many
HEP detectors to other scientific fields, and to government, medical, and industrial applications, it is
becoming increasingly advantageous for universities and laboratories to develop collaborations with industry 
for detector development, fabrication, and commercialization.  The experience of industry with support,
lifecycle, and supply chain management can help mitigate some of the challenges of supporting experiments 
throughout their lifetimes and bring down the total cost of ownership.

We have identified several ways to develop stronger ties between the national laboratories,  
universities, and industry that could provide a backbone for future industrial detector development work. 

One of these is to enhance the communication between the HEP Detector R\&D community and both small and 
large industries that have interests in HEP detector development.  Another is to invite companies from
various industries to future annual CPAD workshops to provide the industrial community with an overview of 
current and future detector development projects.  It is crucial to involve industry at an early stage of an 
experimental program to help align development plans with experimental needs.  Finally, forums at these 
workshops for industries to give presentations on their interests and capabilities would provide 
university and laboratory detector development personnel an overview of the capabilities and interests 
of the industrial community.  

Universities that hope to use engineering, technical, and infrastructure resources at national 
laboratories are often deterred by not being able to charge tasks to an established project.  
A possible solution is for laboratories to set aside a small generic R\&D fund dedicated to funding 
outside users wishing to use laboratory infrastructure as part of a detector R\&D project.  Use of these
funds could be controlled by the laboratory point of contact or designee.

The DOE SBIR/STTR program provides grants to small industries, often in collaborations with universities and 
laboratories, to carry out work of mutual interest.  Cooperative Research and Development Agreements
(CRADAs) are another important tool to increase collaborations between national laboratories and universities
and industry.  These can help with confidentiality and commercialization of technologies and are open to
larger companies.  National laboratories and universities are currently working on reducing the administrative
hurdles involved in establishing these agreements.

\subsubsection{Instrumentation pool}

At Fermilab there is a large Physics Research Equipment Pool (PREP) of commercial instrumentation. 
Fermilab has continued to loan this instrumentation to members of the HEP community who have research 
ties with Fermilab.  Because of financial constraints, this program no longer has the necessary personnel 
to adequately maintain the existing instrumentation pool and to carry out the administrative overhead 
required to maintain an effective loan program.  In order to use this instrumentation in a maximally efficient 
way in the U.S. HEP research community, Fermilab has proposed that the PREP loan program be expanded to the 
entire U.S. HEP research community.  This would require a modest amount of DOE support.  
Furthermore, this pool might serve as a way for companies to expose their products to a large user community.  
It is not clear whether a large, national instrumentation pool is cost-effective for the U.S. HEP research community, 
but it might provide a way for universities and laboratories to collaborate effectively in the use of instrumentation 
resources.

\subsubsection{Synergies between HEP and Nuclear Physics}

Future experiments in nuclear and particle physics will require new instrumentation in order to carry out 
fundamental measurements of QCD, electroweak interactions, studies of fundamental symmetries, understanding the
properties of neutrinos and the spin properties of nucleons, and a variety of other physics topics.  Much of the
instrumentation to accomplish this is common to both nuclear and particle physics and requires the development
of new technologies. Both the nuclear and HEP communities have ongoing 
programs to develop new detector technologies.  Both communities would benefit from identifying areas of mutual interest
and working together to achieve better performance and identify the most cost-effective detectors for future 
experiments, particularly in the development areas described below.

Mutually beneficial developments related to calorimetry include (1) dense, compact electromagnetic calorimeters, 
(2) new techniques for compensating hadronic calorimeters including dual readout calorimetry, (3) new crystal 
calorimeters, (4) new light collection schemes for fiber calorimeters, (5) new and improved silicon photomultipliers 
(SiPMs), particularly extending their dynamic range and improving their photon detection efficiency and 
radiation hardness, and (6) readout systems for large numbers of SiPMs.  

Mutually beneficial developments related to particle tracking include (1) large area gas electron multiplier (GEM) 
detectors, (2) TPCs with GEM readout and short drift GEM detectors, (3) new readout structures for GEM detectors, 
(4) low-mass GEM detectors, (5) TPCs with combined particle ID capability, and (6) low-mass, monolithic, 
active pixel sensors (MAPS).

Mutually beneficial developments related to particle identification include 
(1) Cherenkov ring imaging detectors with GEM readout, 
(2) DIRC detectors, 
(3) large area, high-resolution ($<$ 10 ps) time-of-flight detectors, 
(4) aerogel counters, and 
(5) techniques for muon detection.

Finally, mutually beneficial developments related to readout electronics and triggering include 
(1) ASIC development for micropattern detector readout, 
(2) ASIC development for silicon photomultiplier readout, 
(3) modular readout systems for micro-pattern gas detectors (MPGDs) and SiPMs, and 
(4) fast triggering and large scale readout systems.  

\section{Leadership in detector R\&D}
\label{sec:instrum-leader}

Leadership requires both vision and the ability to execute that vision.  In instrumentation that vision 
must be both scientific and technical.   We need scientists who not only understand the physics of Higgs 
boson decays, but also understand the physics of sensors, materials that 
can be used for supports, ways to interconnect the sensor arrays, the limitations of each, 
and the in-depth knowledge of the physics underlying the instrument to be able to improve its 
science reach.  
These scientists in turn require 
access to infrastructure, engineers, and technicians to turn visions into prototypes and working detectors.  
These resources might be at universities, national laboratories, or industry, but they must be made available to 
scientists with good ideas. A goal of the U.S. instrumentation program must be to enable the execution of 
visionary ideas as well as to support incremental development and system engineering.

Over the years, the U.S. has led many significant detector R\&D efforts in HEP. 
Examples include the invention of the TPC, the development of transition edge sensors for phonon detection, 
the development of CCDs as detectors for cosmology, and the development of liquid argon calorimetry. 
A significant fraction of the overall HEP detector R\&D activities is now being carried out in 
Europe, Japan, and China.  The U.S. has no equivalent to the CERN RD programs or the EU Framework 
initiatives that can address significant technical challenges in a systematic way.

Another factor which contributes to the loss of detector R\&D activity in the U.S. is the 
unwillingness of most U.S. graduate programs to grant a Ph.D. to HEP students who develop 
hardware instead of carrying out an analysis of experimental data.  
The multi-year design and construction time required for detector systems in 
large accelerator experiments makes it impossible for graduate students to take on major hardware roles
in large experiments and graduate in a timely manner.  
In addition, new Ph.D.s with primarily instrumentation development 
backgrounds typically have difficulty getting U.S. university positions.  
To address these problems, incentives must be 
found to encourage younger physicists to become more proficient in hardware development.  Summer programs have been 
organized by national laboratories with HEP programs to give graduate students hands-on experience 
with instrumentation. 

The U.S. maintains clear leadership roles in a number of detector R\&D projects of importance to HEP.  
In recent years the U.S. has developed 3-D sensors, applications of 3-D circuits, water-based scintillators, 
large area photodetectors, cryogenic germanium and silicon detectors with phonon and charge readout, dual readout of 
calorimeters, and deep depletion CCDs.  The U.S. also leads in the development of cryogenic superconducting 
detectors. These developments have been based on the strong interdisciplinary culture among universities 
and national laboratories with utilization of centers of excellence in each.  In the longer term these resources 
can provide a continued focus for innovation and leadership.

\section{Conclusions}
\label{sec:instrum-conclusions}
 
\vspace{0.5cm}  


Modern particle physics experiments and high-energy accelerators put increasingly extraordinary 
demands on the experimental tools employed. 
In addition, the development 
of state-of-the-art detectors is broadening beyond experiments at the highest energy accelerators 
to include lower-energy, high-intensity and ultra-low-background experiments, very large volume 
experiments to study very rare processes, and experiments that study fundamental properties of the 
cosmos.

The field of particle physics finds itself with a panoply of possible large-scale physics experiments, 
each of which has the potential to provide fundamental insight into the structure of space, time, and 
matter.  These experiments are exceedingly difficult technically, and many of them currently
deploy technologies that are expensive, limit physics capabilities, or both. Furthermore, multiple 
approaches are  often required to explore and establish new scientific principles.  New technologies 
must be developed to provide the answers to these fundamental science questions.  
 
Focusing only on scaling up existing technologies to larger experiments, or carrying out 
detector R\&D only when it is needed, is tempting in tight budgetary times, but this approach 
will not serve the field well in the long term.   A development program of new approaches 
and innovation will be necessary to make future experiments feasible and economically viable. 
The demands of the future particle physics science program will require an investment in a 
detector R\&D program over intermediate and long time frames to develop new tools and 
technologies that are both cost-effective and have an enhanced physics reach.  Such an investment will 
enable the field to carry out flagship domestic experimental research and have leadership roles in 
off-shore experimental projects.  New, transformative detection capabilities will ensure 
an affordable and healthy experimental particle physics research program in the future. 
The major challenge for the Instrumentation Frontier is, therefore, to design a program that will 
enable the U.S. to maintain scientific leadership in many key areas of a broad international
experimental program in particle physics.

Ideally, detector development should take place at various levels of risk and maturation time 
scales. Detector R\&D carried out within existing experiments is, by necessity, project-driven 
and provides low-risk, relatively incremental improvements to existing technologies.  Detector 
R\&D that is motivated by common needs among various experiments is generally longer-term, can 
be higher risk, and adds value to multiple experimental areas at the same time. This type of 
R\&D can lead to incremental or significant improvements in cost reduction, scientific reach, 
or both.  Finally, long-term detector R\&D leading to transformative changes in cost 
reduction, increase in scientific reach or both, across a significant part of the experimental 
program is the kind of high-risk, high-reward detector R\&D which has the potential to lead to 
scientific breakthroughs.  Underpinning all these R\&D efforts is the urgent need for training the 
next generation of instrumentation experts, without which there can be no long-term future. 


Major technological advances based on a better understanding of the underlying science are 
also occurring in other scientific disciplines such as materials science, photonics and nanotechnology.  
Many of these advances have the potential to lead to transformational new technologies for 
particle physics detectors.  Technology advances in other experimental 
scientific disciplines that could contribute to opportunities for innovation and developing 
transformative technologies for particle physics must be exploited.   

A healthy national instrumentation program must, therefore, provide a balance between 
evolutionary and revolutionary detector development while training the next generation of experts. 
We recommend that this national program accomplish the following goals:  
\begin{enumerate}
\item Dedicate an appropriate fraction of the HEP budget to supporting detector R\&D with
clearly identified areas of detector development based on the strengths of the HEP community.
\item Achieve an appropriate balance between evolutionary and revolutionary 
detector R\&D, i.e., an appropriate ``portfolio of risk'', and build expertise in new and 
innovative technologies that can be applied to the design and construction of novel, 
cost-effective particle physics detectors.
\item Develop a process for optimizing the use of existing university, national laboratory, 
and industry resources, for growing and retaining local technical expertise at 
universities and laboratories, and for identifying incentives and mechanisms for 
improving detector R\&D collaborations and equipment sharing among universities, national 
laboratories and industry.
\item  Create opportunities for attracting and providing careers for high-energy physicists with interest in,
and outstanding capability for, innovative detector design and development to develop a
community of experimental physicists with the background and skills needed to design and 
build future generations of high-energy physics experiments.
\item Provide mechanisms for identifying and transferring appropriate technologies developed in
other scientific disciplines to high-energy physics and transfer applicable technologies 
developed in high-energy physics to other science disciplines, such as nuclear physics, 
basic energy sciences, and related branches of science, medicine, and national security. 
\end{enumerate}

The Snowmass process, as summarized in this report, provided broad input and guidance from the 
HEP community in identifying the crucial elements of a national technology and 
instrumentation program.  This information will be used by CPAD, the Coordinating Panel for Advanced 
Detectors created by a previous DPF Task Force on Instrumentation, whose mission is to advocate this 
program, promote its merits, and provide venues for regular presentation of results.   CPAD will 
also bring different scientific communities together to make HEP aware of 
developments elsewhere and provide information about instrumentation needs and activities
to facilitate coordination between the funding agencies and the 
HEP instrumentation community. 

An outstanding U.S. instrumentation program requires a variety of enablers, some of which
already exist, and some of which need to be developed.   
The U.S. has a unique resource embodied in its national laboratories, particularly with regard to 
extensive technical infrastructure.  Multidisciplinary laboratories have broad scientific breadth 
and the potential to provide cross-fertilization of technologies from nuclear physics, basic energy science, 
materials research, engineering, chemistry, and computer science.  Many national laboratories have close 
links to top-ranked universities, which are also centers for multi-disciplinary innovation and ideas but 
which generally cannot maintain the level of technical infrastructure available at the laboratories.  By 
combining the intellectual and manpower capabilities of universities with the resources of the national 
laboratories and the product development capabilities of industry, the U.S. can continue to confront and 
overcome many of the technological challenges future particle physics experiments will face. 

Test beams, irradiation facilities, and centers of technical excellence are also program enablers.  
The U.S. has several test beams and irradiation facilities as well as centers of 
technical excellence and expertise, both at universities and laboratories. 
These facilities are a critical component of instrumentation work, and there are  
Snowmass white papers describing them~\cite{testbeams}. However, since the Snowmass process did not 
complete the identification of all the available facilities in the U.S. which support instrumentation 
development, CPAD will need to complete this work.  Adequate support for these facilities is essential 
for a healthy detector R\&D program. 


A crucial enabling element for an instrumentation program is the development of expert
physicist manpower.  Investing in younger physicists who work on instrumentation is critical
for a long-term, viable particle physics program.  Doctorate degrees are awarded in 
instrumentation in the U.S. only under exceptional circumstances, and early careers in
instrumentation are strongly discouraged given the emphasis on analysis to obtain faculty
positions.  A major challenge for CPAD will be finding ways to develop and retain outstanding 
instrumentation physicists within the U.S. HEP community.

We should identify mutually beneficial
collaborations between HEP, other scientific fields, and industry.  The increased complexity of 
modern research experiments and the applications of many HEP detectors to other scientific fields and to 
government, medical, and industrial applications, makes it advantageous for universities and laboratories 
to develop collaborations with industry for detector development, fabrication, and commercialization. 
One mechanism to increase the communication between the particle physics community and both small and 
large industries is to invite these industries to future annual CPAD workshops to provide the industrial 
community with an overview of current and future detector development projects. Industrial forums at 
these workshops would provide the university and laboratory detector development community 
an overview of the capabilities and interests of industry.  It is expected that this 
venue could also lead to a better utilization of the DOE SBIR/STTR and CRADA programs.

This report describes some of the overlaps in the instrumentation developed by particle physics and the 
instrumentation needs in other areas of science.  This holds especially true for the instrumentation needs for 
nuclear physics. All science communities have ongoing detector development programs to develop these technologies, 
and each would benefit from identifying areas of mutual interest and working together to achieve better 
performance and the most cost effective detectors for future experiments. CPAD could play a leading 
role at facilitating this dialogue.

Detectors will need various combinations of finer granularity, higher speed, larger sensitive detection volumes, 
and large area readouts to carry out experiments at higher energies and luminosities, to make more precise and 
inclusive 
measurements, and to search for rare processes.  From the overview given in this report the following strategic 
instrumentation areas can be identified: ASICs, calorimetry, high speed trigger and data acquisition, large volume 
detectors, photodetectors, pixelated sensors, and power and mechanics (see Table~\ref{themes}).  
These areas all focus on the major technological 
barriers that stand in the way of reaching our science goals.  Some of them have the potential to deliver very 
cost-effective instrumentation methods and provide breakthrough technology.  The choice of these areas is guided 
by their physics impact and existing strengths and capabilities in the U.S..  Consideration was given to the 
technology's usefulness to other branches of science.  
Although some of the research goals described in this report may seem very 
challenging, it is likely that many, if not all, can be realized with a dedicated instrumentation effort. 
This selection of strategic themes also illustrates the breadth of their applications in HEP.  

Providing additional support beyond the existing directed and generic R\&D funding to 
initiate a set of \lq\lq grand challenges\rq\rq\,to encourage long term, more speculative,
but potentially higher-impact R\&D in instrumentation would be another enabler
for transformative innovation.  Once identified, such grand challenges would be effective 
in focusing the creative power of the HEP community on problems that have the potential for large 
payoff.  Areas where existing technologies do not provide sufficient physics performance or
would be cost-prohibitive for meeting the goals of future experiments are good candidates for 
new initiatives.  These grand challenges should be issued nationally, and cross-disciplinary 
collaboration should be strongly encouraged.  This would allow the program to take advantage of 
the tremendous advances that have been made in areas of science outside 
the field of particle physics that could prove very valuable for the development of future 
instrumentation.  Because of the cross-disciplinary aspect, close collaboration between universities 
and national laboratories will be key.  Industry should be involved where appropriate.  Funding would 
be subject to proposal review, but should be at a substantial level for a period of at least three years.  
CPAD should be involved in identifying the set of grand challenges that would define the program. 

\begin{table}[htdp]
\caption{Strategic themes in instrumentation with areas of applicability.}
\begin{center}
\begin{tabular}{|l|l|c|c|c||c|c|c|}
\hline
\bf{Instrumentation area} & \bf{Technology examples} & \rotatebox{90}{Energy F. } &  \rotatebox{90}{Intensity F. } 
&  \rotatebox{90}{Cosmic F. }  &  \rotatebox{90}{Nucl. Phys.  }  &  \rotatebox{90}{BES }  &  \rotatebox{90}{Applied }  \\
\hline
Pixelization and tracking & MAPS, 3-D silicon & X & X & X & X & X & X \\
\hline
Mechanics and power & Carbon foams, thin silicon &  X & X & X & X & X & X \\
 & Power delivery  &   &  &  &  &  & \\
\hline
Calorimetry & Crystal EM and hadronic & X  & X &  & X & X &  \\
 & Compensating, PFA  &   &  &  &  &  & \\
\hline
Trigger and DAQ & TCA, high-speed optical links &  X & X & X & X & X & X \\
\hline
Photodetectors & LAPPD, SiPM &  X & X & X & X & X & X \\
\hline
Large volume detectors & Neutron detectors, photodetectors &   & X & X & X &  & X \\
 & Low background materials  &   &  &  &  &  & \\
\hline
ASICs, custom electronics &  Cold, fast electronics, 3-D  &  X & X & X & X & X & X \\
 & Waveform digitization  &   &  &  &  &  & \\
\hline
\end{tabular}
\end{center}
\label{themes}
\end{table}

In particle physics we explore extreme realms of space, time, and energy. 
These explorations require unique instruments.  Developing these instruments requires vision 
and support, both financial and scientific.  The pace of innovation has not yet slowed, but 
the technical and fiscal challenges have increased with the scale of experiments.  
In this report we have reviewed the instrumentation needs in the Energy, Intensity and Cosmic Frontiers.  
We have identified significant opportunities and synergies for instrumentation 
development.  We have argued for programs to support the development of critical and strategic 
technologies, and argued for renewed emphasis on the training and support of young scientists.  
We have reviewed technologies, and identified areas of particular importance and promise.  
We have also argued that a program to address strategic challenges would provide a missing 
component and better prepare the field for the future.  We hope that CPAD can 
provide the future support to the community and advise the agencies, which will enable 
particle physics to realize its vision.  A stably and adequately funded generic instrumentation 
program with a balanced portfolio of risk, opportunities for young physicists, and a dedicated
portion of the HEP budget is essential.  It will ensure that the field invests in its future, and 
establishes a foundation for a competitive, healthy program for the long term.


\begin{thebibliography}{99}



\bibitem{Dyson}
    ``Imagined Worlds'', F.~Dyson, Harvard University Press, 1995; 
    ``The Sun, the Genome, the Internet. Tools of scientific revolutions'',
    F.~Dyson, Oxford University Press, 1997. 

\bibitem{PCAST}
    Report to the President ``Transformation 
    and Opportunity: The Future of the U.S. Research Enterprise'', President's Council of Advisers 
    on Science and Technology, Nov. 2012; www.whitehouse.gov/ostp/pcast.  

\bibitem{DPF-TF}
    Report to the DPF ``Instrumentation in Particle Physics'', October 2011; 
    http://www.hep.anl.gov/cpad/docs/dpf\_report\_v11.pdf. 

\bibitem{CMS:2006}
    Bayatian, G L,
   CERN-LHCC-2006-001; CMS-TDR-8-1.

\bibitem{ATLAS:1999uwa} 
   CERN,
  CERN-LHCC-99-14.

\bibitem{Brooijmans:2013gba} 
  G.~Brooijmans, H.~Evans and A.~Seiden,
  arXiv:1307.5769 [physics.ins-det].

\bibitem{Smith:2013oka} 
  W.~H.~Smith,
  arXiv:1307.0706 [physics.ins-det].


\bibitem{ILC:TDRv3}
   C. Adolphsen {\it et al.}
   arXiv:1306.6328.

\bibitem{CLIC:LCS2012}
   D. Dannheim {\it et al.}
   arXiv:1208.1402.

\bibitem{TLEP:2013}
  The TLEP Design Study Group Home Page:  http://tlep.web.cern.ch

\bibitem{ILC:TDRv4}
   Ties Behnke, et. al.,
   arXiv:1306.6329.

\bibitem{CLIC:CDR2012}
   L. Linssen {\it et al.}
  arXiv:1202.5940.

\bibitem{CALICE:2012}
  C. Adloff {\it et al.}
  arXiv:1212.5127.

\bibitem{DHC:2012}
  K. Francis,
  arXiv:1202.6080

\bibitem{DualReadout:2013}
  N. Akchurin {\it et al.}
  arXiv:1307.5538.


\bibitem{MinosEvans:2013}
  J. Evans,
  arXiv:1307.0721

\bibitem{T2K:2011}
  K. Abe {\it et al.}
  arXiv:1106.1238.

\bibitem{Camilleri:2013oxa} 
  L.~Camilleri [MicroBooNE Collaboration],
  Nucl.\ Phys.\ Proc.\ Suppl.\  {\bf 237-238}, 181 (2013).

\bibitem{Patterson:2012zs} 
  R.~B.~Patterson [NOvA Collaboration],
  Nucl.\ Phys.\ Proc.\ Suppl.\  {\bf 235-236}, 151 (2013)
  [arXiv:1209.0716 [hep-ex]].

\bibitem{LBNE:CD1}
  The LBNE CD1 Review:  http://lbne.fnal.gov/reviews/CD1-FD.shtml.

\bibitem{MINERvA:2006aa} 
  [MINERvA Collaboration],
  FERMILAB-DESIGN-2006-01.

\bibitem{Anderson:2012vc} 
  C.~Anderson, M.~Antonello, B.~Baller, T.~Bolton, C.~Bromberg, F.~Cavanna, E.~Church and D.~Edmunds {\it et al.},
  JINST {\bf 7}, P10019 (2012)
  [arXiv:1205.6747 [physics.ins-det]].

\bibitem{Band:2012rk} 
  H.~R.~Band {\it et al.}
  JINST {\bf 8}, T04001 (2013)
  [arXiv:1206.7082 [physics.ins-det]].

\bibitem{Abe:2012tg} 
  Y.~Abe {\it et al.}  [Double Chooz Collaboration],
  Phys.\ Rev.\ D {\bf 86}, 052008 (2012)
  [arXiv:1207.6632 [hep-ex]].

\bibitem{Kettell} 
  S.H.~Kettell, R.A.~Rameika, R.S.~Tschirhart, 
  ``Intensity Frontier Instrumentation'', arXiv:1309.6704, SNOW13-00162.  

\bibitem{MinFang}
   M.~Yeh {\it et al.}, ``A new water-based liquid scintillator and potential applications'', 
   Nucl. Instrum. Methods. {\bf A660} (2011) 51 - 56.

\bibitem{Farine:2013raa} 
  J.~Farine [EXO Collaboration],
  Nucl.\ Phys.\ Proc.\ Suppl.\  {\bf 235-236}, 255 (2013).

\bibitem{Fukuda:2002uc} 
  Y.~Fukuda {\it et al.}  [Super-Kamiokande Collaboration],
  Nucl.\ Instrum.\ Meth.\ A {\bf 501}, 418 (2003).

\bibitem{Regis:2012sn} 
  C.~Regis {\it et al.}  [Super-Kamiokande Collaboration],
  Phys.\ Rev.\ D {\bf 86}, 012006 (2012)
  [arXiv:1205.6538 [hep-ex]].


\bibitem{Nishiguchi:2011zz} 
  H.~Nishiguchi [MEG Collaboration],
  AIP Conf.\ Proc.\  {\bf 1382}, 239 (2011).

\bibitem{Iwai:2012qya} 
  E.~Iwai [KOTO Collaboration],
  Nucl.\ Phys.\ Proc.\ Suppl.\  {\bf 233}, 279 (2012).

\bibitem{Knoepfel:2013ouy} 
  K.~Knoepfel {\it et al.}  [mu2e Collaboration],
  arXiv:1307.1168 [physics.ins-det].

\bibitem{Worcester:2012rd} 
  E.~T.~Worcester [ORKA Collaboration],
  Nucl.\ Phys.\ Proc.\ Suppl.\  {\bf 233}, 285 (2012)
  [arXiv:1211.4883 [hep-ex]].

\bibitem{g-2} 
  F.~Jegerlehner, A.~Nyffeler, Phys. Rept. 477 (2009) 1. 


\bibitem{LHCb:UpgradeTDR}
    LHCb Collaboration,
   CERN-LHCC-2012-007; LHCb-TDR-12.

\bibitem{Abe:2010sj} 
  T.~Abe [Belle II Collaboration],
  arXiv:1011.0352 [physics.ins-det].

\bibitem{Wang:2007sm} 
  Y.~-F.~Wang,
  eConf C {\bf 070805}, 38 (2007)
  [arXiv:0711.4199 [hep-ex]].

\bibitem{Roney:2013}
  Roney, JM,
  arXiv:1304.2431.


\bibitem{Allekotte:2007sf} 
  I.~Allekotte {\it et al.}  [Pierre Auger Collaboration],
  Nucl.\ Instrum.\ Meth.\ A {\bf 586}, 409 (2008)
  [arXiv:0712.2832 [astro-ph]].

\bibitem{Abraham:2009pm} 
  J.~Abraham {\it et al.}  [Pierre Auger Collaboration],
  Nucl.\ Instrum.\ Meth.\ A {\bf 620}, 227 (2010)
  [arXiv:0907.4282 [astro-ph.IM]].

\bibitem{Kiryluk:2008dm} 
  J.~Kiryluk [IceCube Collaboration],
  arXiv:0806.1717 [astro-ph].

\bibitem{PINGU:2013}
   IceCube [PINGU collaboration]
  arXiv:1306.5846.

\bibitem{ANITA:2013}  
  K. Belov,
  arXiv:1303.2172.


\bibitem{DES:2005}
  The Dark Energy Survey Collaboration,
  arXiv:astro-ph/0510346.

\bibitem{Dawson:2012va} 
  K.~S.~Dawson {\it et al.}  [BOSS Collaboration],
  arXiv:1208.0022 [astro-ph.CO].

\bibitem{Levi:2013}
  MS-DESI: https://indico.fnal.gov/materialDisplay.py?contribId=38\&sessionId=22\&materialId=slides\&confId=6199

\bibitem{Euclid:2012}
  L. Amendola, {\it et al.}
  arXiv:1206.1225.

\bibitem{LSST:2011}
  Z. Ivezic {\it et al.}
  arXiv:0805.2366.

\bibitem{Gigaz:2013}
  D. Marsden {\it et al.}
  arXiv:1307.5066.


\bibitem{VERITAS:1999}
  V.V. Vassiliev, {\it et al.}
  arXiv:astro-ph/9908135.

\bibitem{VERITAS:2011}
  D.B. Kieda,
  arXiv:1110.4360.

\bibitem{HESS:2003}
  U. Schwanke,
  arXiv:astro-ph/0307287.

\bibitem{MAGIC:2005}
  D. Bastieri {\it et al.}
  arXiv:astro-ph/0503534.

\bibitem{HAWC:2013}
  D. Zaborov,
  arXiv:1303.1564.

\bibitem{CTA:2009}
  R.M. Wagner {\it et al.}
  arXiv:0912.3742.


\bibitem{CDMS:2005}
  D. S. Akerib {\it et al.} (CDMS) 2005 {\it Phys. Rev.}{\bf D72} 052009.

\bibitem{CoGeNT:2012}
  C.E. Aalseth {\it et al.}
  arXiv:1208.5737.

\bibitem{DAMIC:2012}
  J. Barreto {\it et al.}
  arXiv:1105.5191.

\bibitem{Xenon100:2012}
  E. Aprile {\it et al.}
  arXiv:1107.2155.

\bibitem{LUX:2012}
  D.S. Akerib {\it et al.}
  arXiv:1211.3788.

\bibitem{Darkside:2011}
  A. Wright,
  arXiv:1109.2979.

\bibitem{MiniCLEAN}
  A. Hime,
  arXiv:1110.1005.

\bibitem{COUPP:2008}
  E. Behnke {\it et al.}
  arXiv:0804.2886.

\bibitem{XENON1T:2012}
  E. Aprile,
  arXiv:1206.6288.

\bibitem{LZ:2011}
  D.C. Malling {\it et al.}
  arXiv:1110.0103.

\bibitem{XAX:2012}
  K. Arisaka {\it et al.}
  arXiv:1107.1295.

\bibitem{COUPP500}
  COUPP500:  http://www-coupp.fnal.gov/public/500kg\ PAC\ Proposal.pdf

\bibitem{DMTPC}
  J.B.R.~Battat, {\it et al.}, ``DMTPC: Dark matter detection with directional sensitivity'', 
   arXiv:1012.3912. 

\bibitem{DRIFT}
  E.~Daw {\it et al.}, ``The DRIFT Dark Matter Experiments'', 
  arXiv:1110.0222.  

\bibitem{vanBibber:2013ssa} 
  K.~van Bibber and G.~Carosi,
  arXiv:1304.7803 [physics.ins-det].


\bibitem{CMB:2013}
  D. H. Weinberg {\it et al.}
  arXiv:1309.0842.

\bibitem{CMB:Bmode}
  D.~Hanson {\it et al.}
  arXiv:1307.5830v1


\bibitem{Hogan:2009}
  Craig J. Hogan, Mark G. Jackson
  arXiv:0812.1285.


\bibitem{sensors}
A. Seiden, M. Artuso {\it et al.}, ``Sensor Compendium'', 
   arXiv:1310.5158,  
FERMILAB-FN-0969-PPD, ANL-HEP-TR-13-51.

\bibitem{MAPS:2006}
  R. Turchetta,
  arXiv:physics/0605238.

\bibitem{Barbero} 
M. Barbero, 
in http://www.bo.infn.it/sminiato/siena13.html.

\bibitem{Mekkaoui}
A. Mekkaoui et. al., 
2013 International Image Sensor Workshop,
http://www.imagesensors.org/Past\%20Workshops/2013\%20Workshop/2013\%20Papers/05-03\_008-mekkaoui\_paper.pdf.

  
\bibitem{Hara:2011hia} 
  K.~Hara, K.~Shinsho, T.~Ishibashi, Y.~Arai, T.~Miyoshi, Y.~Ikemoto, R.~Ichimiya and T.~Tsuboyama {\it et al.},
  IEEE Nucl.\ Sci.\ Symp.\ Conf.\ Rec.\  {\bf 2011}, 1045 (2011).

\bibitem{SOI:2008}
  M. Battaglia {\it et al.}
  arXiv:0811.4540.

\bibitem{Deptuch:2010zz} 
  G.~Deptuch, M.~Demarteau, J.~Hoff, R.~Lipton, A.~Shenai, R.~Yarema and T.~Zimmerman,
  FERMILAB-CONF-10-401-PPD.
  
\bibitem{Yarema:2008zza} 
  R.~Yarema, D.~Christian, M.~Demarteau, G.~Deptuch, J.~Hoff, R.~Lipton, A.~Shenai and M.~Trimpl {\it et al.},
  CERN-2008-008.
  
\bibitem{Parker:1997vy} 
  S.~I.~Parker, C.~J.~Kenney and J.~Segal,
  ICFA Instrum.\ Bull.\  {\bf 14}, 30 (1997).
  

\bibitem{Trischuk:2009}
  W. Trischuk [RD42 Collaboration]
  arXiv:0810.3429.


\bibitem{FastSilicon:2012}
  H. Sadrozinski, ``Exploring charge multiplication for fast timing with silicon sensors'', \\
  https://indico.cern.ch/conferenceOtherViews.py?view=standard\&confId=175330.

\bibitem{Deptuch:2013soa} 
  G.~Deptuch, U.~Heintz, M.~Johnson, C.~Kenney, R.~Lipton, M.~Narian, S.~Parker and A.~Shenai {\it et al.},
  arXiv:1307.4301 [physics.ins-det].

\bibitem{Hohlmann:2013aga} 
  M.~Hohlmann, V.~Polychronakos, A.~White and J.~Yu,
  arXiv:1306.1924 [physics.ins-det].

\bibitem{power}
W. Cooper, C. Haber, D. Lynn,
``Low Mass Support and Cooling for Future Collider Tracking Detectors'',
Whitepaper submitted to the Instrumentation Frontier,  \\
http://www.snowmass2013.org/tiki-index.php?page=Instrumentation+Frontier+Whitepapers.


\bibitem{Dhawanpower}
S.~Dhawan, R. Sumner,R. Khanna, ``Powering Future Particle Physics Detectors,'' Whitepaper submitted to 
the Instrumentation Frontier, \\
 http://www.snowmass2013.org/tiki-index.php?page=Instrumentation+Frontier+Whitepapers.


\bibitem{Hohlmann:3D}
  M. Hohlmann,
  arXiv:1309.0842.


\bibitem{DeGeronimo:2013zxa} 
  G.~De Geronimo, D.~Christian, C.~Bebek, M.~Garcia-Sciveres, H.~Von der Lippe, G.~Haller, 
  A.~A.~Grillo and M.~Newcomer,
  arXiv:1307.3241 [physics.ins-det].


\bibitem{pxemcal}
F. DeJongh {\it et al.}, ``Electromagnetic Calorimetry in Project X Experiments Ð The Project X Physics Study
Summary and R\&D prospects,'' Whitepaper submitted to 
the Instrumentation Frontier,  
http://www.snowmass2013.org/tiki-index.php?page=Instrumentation+Frontier+Whitepapers

\bibitem{Albayrak-Yetkin:2013xga} 
  A.~Albayrak-Yetkin, B.~Bilki, J.~Corso, P.~Debbins, G.~Jennings, V.~Khristenko, A.~Mestvirisvilli 
  and Y.~Onel {\it et al.},
  arXiv:1307.8051 [physics.ins-det].

\bibitem{Albayrak-Yetkin:2013yma} 
  A.~Albayrak-Yetkin, B.~Bilki, J.~Corso, G.~Jennings, A.~Mestvirisvilli, Y.~Onel, I.~Schmidt and C.~Sanzeni 
  {\it et al.},
  arXiv:1307.8376 [physics.ins-det].

\bibitem{zhucal}
Ren-Yuan Zhu, ``The Next Generation of Crystal Detectors,'' Whitepaper submitted to 
the Instrumentation Frontier,  \\
http://www.snowmass2013.org/tiki-index.php?page=Instrumentation+Frontier+Whitepapers, 
FERMILAB-FN-0969-PPD.


\bibitem{Ossy_highres}
  A.~Tremsin and O.~Siegmund, ``Spatial distribution of electron cloud foot- prints from microchannel 
  plates: Measurements and modeling'', Rev. Sci. Instrum. 70, 3282 - 3288 (1999).

\bibitem{timingref}
  J.~Milnes, J.~Howorth, ``Picosecond time response of microchannel plate PMT
  detectors'', Proc. SPIE 5580 (2005) 730--740.

\bibitem{Akatsu:2004mq}
  M.~Akatsu, Y.~Enari, K.~Hayasaka, T.~Hokuue, T.~Iijima, K.~Inami, K.~Itoh,
  Y.~Kawakami, {\it et al.}, ``MCP-PMT timing property for single photons'', Nucl.
  Instrum. Methods {\bf A528} (2004) 763--775.

\bibitem{InamiRef}
  K.~Inami, N.~Kishimoto, Y.~Enari, M.~Nagamine, T.~Ohshima, 
  ``A 5-ps tof-counter with an MCP-PMT'', Nucl. Instrum. Methods {\bf A560} (2006) 303--308.

\bibitem{VavraTOF}  
  J. Va'vra {\it et al.}, ``Beam test of a Time-of-Flight detector prototype'', 
  Nucl. Instrum. Methods {\bf A606} (2009) 404.
 
\bibitem{RonzhinTOF} 
  A. Ronzhin {\it et al.}, ``Development of a 10~ps level time of flight system in the 
  Fermilab Test Beam Facility'', Nucl. Instrum. Methods {\bf A623} (2010) 931.

 
\bibitem{LAPPDneut}
  M.~C.~Sanchez and M.~Wetstein, ``Using Large Area Micro-channel Plate Photodetectors in the 
  Next Generation Water Cherenkov Neutrino Detectors''. 
  Nuclear Physics B Proceedings Supplements, Volume 229, p. 525-525 (2010). 

\bibitem{LAPPDneutWP}
  Z.~Djurcic M.~C.~Sanchez, M.~Wetstein {\it et al.}, Snowmass contribution \\
  http://if-neutrino.fnal.gov/whitepapers/lappd.pdf

\bibitem{VavraFDIRC}
  J. Va'vra {\it et al.}, ``Progress on development of the new FDIRC PID detector'', 
  Nucl. Instrum. Methods {\bf A718} (2013) 541.
 
\bibitem{BellII-TOF}
  K.~Nishimura, ``The time-of-propagation counter for Belle-II'', 
  Nucl. Instrum. Methods {\bf A639} (2011) 177.
 

\bibitem{VavraPID}  
  J. Va'vra {\it et al.}, ``PID techniques: Alternatives to RICH methods'', 
  Nucl. Instrum. Methods {\bf A639} (2011) 193.
 

\bibitem{Xie} 
  J.Q.~Xie, M.~Demarteau, R.~Wagner, ``High Performance Photocathodes based on Molecular 
  Beam Epitaxy Deposition for Next Generation Photo Detectors and Light Sources'' 
  arXiv:1310.2649 [physics.ins-det].

\bibitem{photocathodeworkshop}
See, for example, the first and second workshops on photocathodes at University of Chicago: \\
https://psec.uchicago.edu/workshops/photocathodeConference/ and  \\ 
https://psec.uchicago.edu/workshops/2nd\_photocathode\_conference/ \, . 

\bibitem{LehmanMCPs} 
  A. Lehmann {\it et al.}, ``Systematic studies of MCP-PMTs'', 
  Nucl. Instrum. Methods {\bf A639} (2011) 144.





 
 


 

\bibitem{LAPPDWP}
    E.~Oberla, J.-F.~Genat, H.~Grabas, H.~Frisch, K.~Nishimura, G.~Varner,  
    ``A 15 GSa/s, 1.5 GHz Bandwidth Waveform Digitizing ASIC'', 
    Nuclear Instrum. Methods {\bf A735} (2014) 452
    (arXiv:1309.4397). 

\bibitem{MLAP} 
S. Qian, http://ssrc.inp.nsk.su/conf/AFAD2013/presentations/WG3/SenQian--V2.1.pdf

\bibitem{TOFMRPC}
C. Williams, TOF with MRPC, Proceedings of the Erice Workshop, 2003.

\bibitem{STARMRPC}
Ming Shao et al., STAR MRPC, NIM A 558 (2006) 419.

\bibitem{newMRPC} 
Y.J. Sun et al., New Prototype MRPC with Long Strips, arXiv:0805.2459.






\bibitem{wireless} P.~De~Lurgio, Z.~Djurcic, G.~Drake, R.~Hashemian, A.~Kreps, M.~Oberling, T.~Pearson, H.~Sahoo, 
``A Prototype of Wireless Power and Data Acquisition System for Large Detectors'', 
arxiv:1310.1098 [physics.ins-det].

\bibitem{lbkd}
J. Hall, ``Low Background Materials for Direct Detection of Dark Matter,'' Whitepaper submitted to 
the Instrumentation Frontier, \\ 
http://www.snowmass2013.org/tiki-index.php?page=Instrumentation+Frontier+Whitepapers

\bibitem{seidel}
S. Seidel, ``Irradiation Facilities in New Mexico,'' Whitepaper submitted to 
the Instrumentation Frontier,  \\
http://www.snowmass2013.org/tiki-index.php?page=Instrumentation+Frontier+Whitepapers

\bibitem{testbeams}
 ``Beam Test Facilities'',  ``Fermilab Irradiation Facilities for Instrumentation Research'',  ``Radiation Facilities in New Mexico'',
 in  ``Compendium of Instrumentation Frontier Whitepapers on Technologies for Snowmass 2013'', FERMILAB-FN-0971-PPD.

\end{thebibliography}
\end{document}